\documentclass[floatfix,10pt,aps,prb,twocolumn,showpacs,a4paper]{revtex4-2}

\usepackage{makecell, cellspace}
\usepackage{subfigure}
\usepackage{amsmath}
\usepackage{amssymb}
\usepackage{amsmath}
\usepackage{amssymb}
\usepackage{subfigure}
\usepackage{epstopdf}
\usepackage{makecell, cellspace, caption}
\usepackage{color}
\usepackage{textcomp}
\usepackage{ulem}
\usepackage{bm}
\usepackage{graphicx,rotating,booktabs}
\usepackage{subfigure}
\usepackage{epsfig}
\usepackage{rotating}
\usepackage{amsfonts}
\usepackage{enumerate}
\usepackage{adjustbox}
\usepackage{dcolumn}
\renewcommand{\thetable}{S\arabic{table}}
\usepackage{bm}
\usepackage{graphicx,epsfig}
\renewcommand{\thefigure}{S\arabic{figure}}
\usepackage{times}
\usepackage{appendix}
\usepackage{caption}
\usepackage{float}
\usepackage[pagebackref=false,colorlinks,linkcolor=blue,citecolor=blue,urlcolor=blue]{hyperref}
\usepackage{graphicx,epsfig}
\usepackage{multirow,varwidth}
\usepackage{subfigure}

\usepackage{graphicx}
\usepackage{float}
\begin{document}

\author{Zahra Mosleh}
\email{z.mosleh@ph.iut.ac.ir}
\affiliation{Department of Physics, Isfahan University of Technology, Isfahan 84156-83111, Iran.}
\author{Mojtaba Alaei}
\email{m.alaei@iut.ac.ir}
\affiliation{Department of Physics, Isfahan University of Technology, Isfahan 84156-83111, Iran.}
\title{Benchmarking density functional theory on the prediction of antiferromagnetic transition temperatures}

\begin{abstract}
This study investigates the predictive capabilities of common DFT methods (GGA, GGA+$U$, and GGA+$U$+$V$) for determining the transition temperature of antiferromagnetic insulators.
We utilize a dataset of 29 compounds and derive Heisenberg exchanges based on DFT total energies of different magnetic configurations.    
To obtain exchange parameters within a supercell, we have devised an innovative method that utilizes null space analysis to identify and address the limitations imposed by the supercell on these exchange parameters.
With obtained exchanges, we construct Heisenberg Hamiltonian to compute Transition temperatures using classical Monte Carlo simulations.           
To refine the calculations, we apply linear response theory to compute on-site ($U$) and intersite ($V$) corrections through a self-consistent process. Our findings reveal that GGA significantly overestimates the transition temperature (by ~113\%), while GGA+$U$ underestimates it (by ~53\%). To improve GGA+$U$ results, we propose adjusting the DFT results with the $(S+1)/S$ coefficient to compensate for quantum effects in Monte Carlo simulation, resulting in a reduced error of 44\%.
Additionally, we discover a high Pearson correlation coefficient of approximately 0.92  between the transition temperatures calculated using the GGA+$U$ method and the experimentally determined transition temperatures. Furthermore, we explore the impact of geometry optimization on a subset of samples. Using consistent structures with GGA+U and GGA+U+V theories reduced the error. 
\end{abstract}

\maketitle

\section{Introduction}
Magnetic materials find numerous applications, ranging from energy harvesting to computer memories ~\cite{app1,app2, app3}.
The key to utilizing these materials effectively lies in understanding their magnetic order and transition temperature, as these properties determine their potential applications. Determining these properties requires expensive experiments like Neutron diffraction ~\cite{neutron}, which can be impractical for all compounds created in labs. Therefore, there is a growing need for reliable and computationally efficient ab initio methods to accurately predict magnetic order and transition temperature.

Additionally, computational material scientists have gained the ability to simulate numerous non-existent materials using a combination of ab initio methods, like density functional theory (DFT), and advanced techniques such as machine learning and evolutionary algorithms~\cite{uspex1}. The accurate prediction of magnetic properties for these materials is essential to discover new magnetic substances with desirable characteristics~\cite{Zhang2021}. This significance is particularly pronounced for magnetic materials, as there is a pressing need for expanded databases to explore and identify novel materials using machine-learning approaches effectively.

Theoretical predictions of material properties, including thermodynamics, can be obtained from quantum ab initio calculations. However, these calculations are computationally expensive, making them feasible primarily at zero temperature. To address this issue, various strategies have been developed. One practical approach involves mapping essential material features onto a simpler physical model. In the context of magnetic materials and magnetic energy models, a key feature is the magnetic interactions between atomic magnetic moments such as Heisenberg exchanges. The strength of these interactions can be determined through ab initio calculations using different strategies. Subsequently, these interactions are used to construct a magnetic model. To obtain macroscopic properties, such as thermodynamic properties like phase transition temperature and Curie-Weiss temperature of the magnetic material, classical spin Monte Carlo (MC) simulations are employed for the magnetic model.

Most magnetic materials contain atoms with d or f orbitals in their valence states, resulting in strong electron-electron correlations that pose challenges for most ab initio methods. Some sophisticated techniques, such as dynamical mean-field theory (DMFT)~\cite{DMFT} and continuum quantum Monte Carlo~\cite{QMC1,QMC2}, can address these challenges with a high level of accuracy. However, these methods are computationally expensive and practical only in specific cases. A more practical and cost-effective solution to improve electron-electron interactions in DFT is to incorporate the Hubbard correction, known as DFT+$U$~\cite{DFTU1, DFTU2,widita2021revisiting}. The Hubbard $U$ parameter serves as a regulatory factor and can be estimated using various approaches, such as linear response theory. 

This paper presents a systematic study focused on assessing the effectiveness of linear response theory in predicting appropriate $U$ parameters for accurately calculating magnetic interactions in insulating antiferromagnetic materials. To do this, we specifically select magnetic materials for which experimental data, such as phase transition temperature, are available. Additionally, we explore the effects of geometrical optimization and the calculation of $U$ through a self-consistent process. To derive thermodynamic properties, we employ Monte Carlo simulations (MC) based on magnetic interactions obtained from DFT+$U$ calculations. Finally, we assess the accuracy of the $U$ parameters by comparing transition temperatures obtained through MC simulations with experimental data. Along with the DFT+$U$ benchmark, we provide a benchmark for generalized gradient approximation (GGA) of exchange-correlation functional in DFT in predicting transition temperature.

\section{Materials and Computational Methods}

\subsection{Materials}
In this study, we choose 29 different antiferromagnetic crystals, 
from well-known transition monoxides such as MnO and NiO to complicated compounds such as LiMnPO$_4$. 
We try to have a variety of crystal symmetries in our choice. 
We restrict our selections among materials with 3d valance magnetic atoms to avoid spin-orbit effects. 
Because DFT, i.e., GGA and local density approximation (LDA), underestimates the magnetic moment of some itinerant magnetic materials~\cite{Sharma2018}, 
we limit the candidate to only insulator antiferromagnetic systems. 
The chemical formulas of our compounds are: BiFeO$_3$\cite{selbach2007synthesis},  CoWO$_4$\cite{ahmadi2016synthesis},  Cr$_2$O$_3$\cite{larbi2017structural},  Cr$_2$TeO$_6$\cite{kunnmann1968magnetic},  Cr$_2$WO$_6$\cite{kunnmann1968magnetic},  CrCl$_2$\cite{hagiwara1995magnetic},  
Fe$_2$O$_3$\cite{navale2013synthesis},  Fe$_2$TeO$_6$\cite{kunnmann1968magnetic},  KMnSb\cite{schucht1999magnetischen},  KNiPO$_4$\cite{lujan1994magnetic},  La$_2$NiO$_4$\cite{zhou2009synthesis},  
LaFeO$_3$\cite{idrees2011anomalous},  Li$_2$MnO$_3$\cite{strobel1988crystallographic},  LiCoPO$_4$\cite{newnham1965crystallographic},  LiMnO$_2$\cite{galakhov2000electronic},  LiMnPO$_4$\cite{geller1960refinement},  LiNiPO$_4$\cite{abrahams1993structure},  
MnF$_2$\cite{li2009solvothermal},  MnO\cite{tran2006hybrid},  MnS\cite{clark2021inelastic},  MnSe\cite{peng2002selective},  MnTe\cite{reig2001growth},  MnWO$_4$\cite{saranya2012synthesis},  NiBr$_2$\cite{day1976optical},  NiF$_2$\cite{costa1993charge},  NiO\cite{tran2006hybrid},  NiWO$_4$\cite{keeling1957structure},YFeO$_3$\cite{racu2015direct}, and YVO$_3$\cite{guzman2013synthesis}. 

\subsection{Computational methods}
\subsubsection{Ab initio computational details}
The total energy calculations are performed using density functional theory in two different codes: 
the plane-wave pseudopotential Quantum-Espresso (QE) package~\cite{QE} and the full-potential local-orbital (FPLO) code~\cite{FPLO}. 
In QE, we utilize GBRV ultra-soft pseudopotentials~\cite{GBRV}. We use the experimental crystal structure parameters for all cases.
We adopt the generalized gradient approximation (GGA) in the Perdew-Burk-Ernzerhoff form (PBE) 
for the exchange-correlation potential. 
To expand the wave function and charge density in the plane wave, we set 40 Ry and 400 Ry cut-offs for all compounds, respectively.
To sample the Brillouin zone (BZ), we employ the Monkhorst-Pack scheme ~\cite{MK} with a mesh spacing of 0.2 $\frac{1}{A^\circ}$.

Due to the known underestimation of electron-electron Coulomb interactions in GGA, 
we apply the Hubbard $U$ correction method, commonly called GGA+$U$ (or LDA+$U$). 
We estimate the self-consistent Hubbard $U$ parameter with a precision of about 0.01 eV 
using linear response theory~\cite{LR} through density functional perturbation theory (DFPT)~\cite{LR_DFPT1,LR_DFPT2}.

We employ the Bader charge analysis code~\cite{Bader2006,Bader2007,Bader2009,Bader2011} for charge distribution, 
magnetic moment analysis, and determining the percentage of atoms ionizations in each compound.
\subsubsection{Deriving Heisenberg exchange parameters}
To obtain Heisenberg exchange interactions, we map the spin-polarized DFT total energy of different magnetic configurations into the following Heisenberg Hamiltonian:
\begin{equation}
\label{eq:eq1}
H=-\frac{1}{2}  \sum_{i,j} J_{ij} \hat S_i .\hat S_j
\end{equation}
\noindent
$J_{ij}$ indicates the strength of Heisenberg exchange interaction (Heisenberg exchange parameters) 
between  $i$th and $j$th sites, and $\hat S_i$ denotes the unit vector of magnetic moment direction on the lattice site  $i$. 
We need only consider the colinear spin configurations to derive the Heisenberg exchange. 
So instead of assigning the direction of the magnetic moment by vectors, 
we only need to assign them by $\pm 1$ (i.e., $\hat S_i=\pm 1 $). 

To calculate the exchange parameters up to the $n$th nearest neighbor, 
the distance of the $n$th nearest neighbor should fit within the crystal cell. 
Therefore, we extend the primitive cell to an appropriate supercell for this purpose.
However, determining the farthest neighbor for which we can calculate the exchange parameter 
should not solely rely on comparing the distance of neighbors and the size of the supercell. 
An additional criterion must be considered due to the periodic boundary conditions. Below, we elaborate on this crucial criterion.

Given a magnetic configuration, for instance, the $k$th configuration, 
where the magnetic moment directions are specified with specific values 
(e.g., $\hat S_1 = 1$, $\hat S_2 = -1$, $\hat S_3 = -1$, and so on), 
the Heisenberg Hamiltonian for this configuration can be expressed as follows:
\begin{equation}
\label{eq:eq2}
E_k=\sum_{i}^m \alpha_{ki} J_i + c
\end{equation}
In this equation, $E_k$ represents the total energy of the $k$th magnetic configuration obtained from DFT, 
$J_i$ denotes the Heisenberg exchange parameter for the $i$th nearest neighbor, $c$ is a constant, 
and $\alpha_{ki}$ indicates the coefficient corresponding to the $i$th nearest neighbor for this particular configuration. 
The values of $\alpha_{ki}$ are determined by the specific magnetic moment directions 
(e.g., $\hat S_1$, $\hat S_2$, $\hat S_3$, and so on) for the $k$th configuration.
Each magnetic configuration (e.g., $k$) will have its set of coefficients $\alpha_{ki}$ 
corresponding to the different nearest neighbors, leading to a matrix of coefficients. 
Analyzing the null space~\cite{linearAG} of this coefficient matrix can provide insights 
into the dependencies between the coefficients for different nearest neighbors, 
allowing us to determine the farthest neighbor for which we are permitted to calculate the Heisenberg exchange parameters. 
We explain this analysis in more detail in the Appendix
\ref{app:B}.

To obtain the Heisenberg exchanges up to the $n$th nearest neighbor,
theoretically, we only need a minimum of $n+1$ distinct magnetic configurations. 
However, in practice, additional complexity arises from the induced magnetic moments of anions, such as oxygen atoms.
Consequently, using more magnetic configurations 
and determining the Heisenberg exchanges through the least squares method is advisable, 
as recommended in our previous work~\cite{Alaei2023}. 
Therefore, to account for these additional complexities and ensure more accurate results, 
we use a number of magnetic configurations around three times greater than the minimum required number.
\subsubsection{Monte Carlo simulation}
To calculate the antiferromagnetic transition temperature (i.e., N\'eel temperature) of all compounds, 
we perform classical Monte Carlo (MC) simulations using the replica exchange method 
as implemented in the Esfahan Spin Simulation package (ESpinS) 
as an open-source classical spin MC software package~\cite{ESpinS}. 
We choose the supercell for each compound in a way that the supercell contains around 2000 sites. 
For each replica, at each temperature, 50000 Monte Carlo steps (MCs) per spin are considered for the thermal equilibrium and data collection, respectively. Measurements are done after skipping every 5 MCs to lower the correlation between the data.
\section{Results and discussion}
\subsection{Accuracy of GBRV pseudopotentials for Heisenberg exchanges}
\begin{figure}[H]
	\includegraphics[scale=0.650]{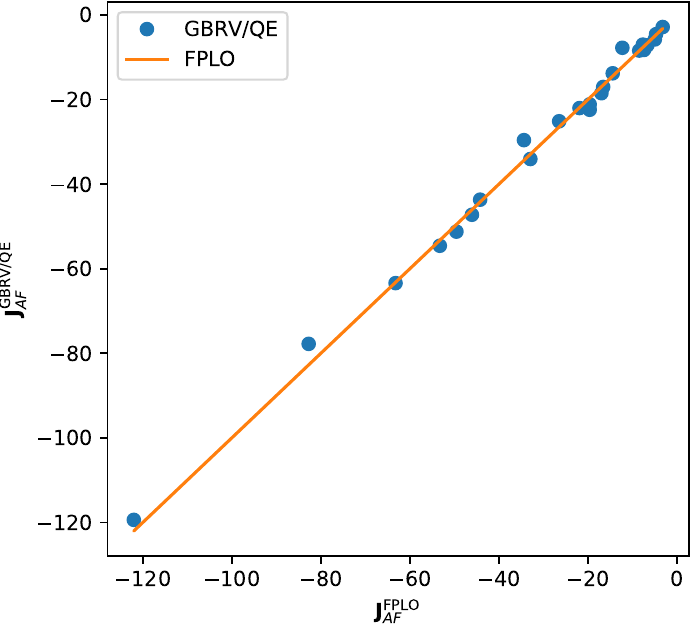}
	\caption{Comparison between Full potential method using FPLO code and GBRV pseudopotentials using QE code. We plot strongest antiferromagnetic exchange for FPLO and GBRV. The line and circles indicate FPLO and QE results, respectively. In the plot, the results of La2NiO4, LiCoPO4, and CoWO4 are absent since FPLO calculations for some of the magnetic configurations of these compounds do not converge.}
	\label{fig:qe-fplo}
\end{figure}

One of the critical aspects in ab initio calculations is the reliability of pseudopotentials, 
especially when dealing with 3d transition metals. 
This is particularly crucial due to the impact of semi-core states and the highly localized nature of 3$d$ electrons~\cite{pseudo1, pseudo2}. 
While GBRV pseudopotentials have been thoroughly examined, their accuracy in predicting Heisenberg exchange interactions lacks a benchmark.
To address this, we conduct a comparison between full-potential FPLO results 
and QE GBRV pseudopotentials within the GGA exchange-correlation functional. 
For this comparison, we use 12 different magnetic configurations for each compound.
Figure~\ref{fig:qe-fplo} indicates the strongest antiferromagnetic exchange values for both FPLO and GBRV. 
Notably, there is a high degree of consistency between the all-electron FPLO and GBRV/QE methods, 
with an average difference of around 1.4 meV. 
Further detailed results can be found in Table S1, S2, S3, and S4 of the Supplemental Material.
\subsection{Transition Temperatures}
To predict the transition temperatures for all samples, 
we employ the Heisenberg exchange parameters obtained from both GGA and GGA+$U$ calculations. 
We use these parameters in Monte Carlo (MC) simulations, which allow us to determine the transition temperatures based on the peak of the magnetic-specific heat.
In the the Supplemental Material , we provide comprehensive details of the Heisenberg exchange parameters and transition temperatures in Tables S5, S6, S7 and S8 for all compounds using both GGA and GGA+$U$ methods. 
\begin{figure}[]
	\includegraphics[scale=0.650]{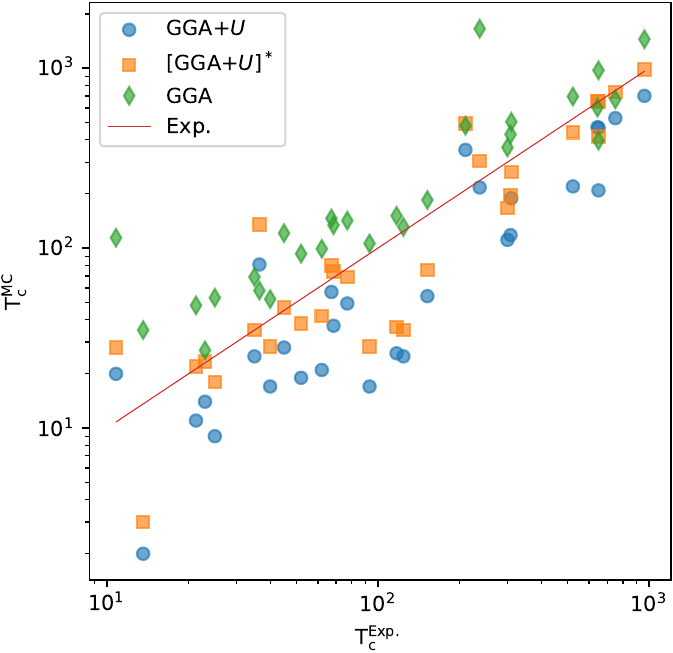}
	\caption{The plot compares the experimental transition temperature of 29 compounds with GGA and GGA+U. We compute the transition temperatures for GGA and GGA+U using monte carlo simulation. The [GGA+$U$]* data show implementing $(S+1)/S$ correction to GGA+$U$.}
	\label{fig:Tc}
\end{figure}
Figure~\ref{fig:Tc} presents a comparison between the experimental transition temperatures, $T_C^{\mathrm{Exp.}}$ 
(depicted by the solid line) and the results obtained from GGA and GGA+$U$ calculations, indicating with $T_C^{\mathrm{MC}}$ axis.
The GGA method significantly overestimates the transition temperatures, with a mean absolute percentage error (MAPE) of 113\%. 
However, in three cases (La$_2$NiO$_4$, LaFeO$_3$, and YFeO$_3$), GGA underestimates the transition temperature.
On the other hand, the GGA+$U$ approach generally underestimates transition temperatures, 
except for three compounds (Fe$_2$TeO$_6$, Li$_2$MnO$_3$, and MnWO$_4$) where they are overestimated. 
The GGA+$U$ method reduces the MAPE in estimating transition temperatures to 53\%.
Despite these errors, We observe Pearson correlation coefficients of 0.74 and 0.92 for the transition temperatures of GGA and GGA+U, 
respectively, when compared to the experimental values. 
These findings can be valuable for future research, especially in the context of machine learning applications.

Due to the classical nature of our MC simulations, in our previous work~\cite{Alaei2023}, 
we introduced a correction factor of $(S+1)/S$ as follows:
\begin{equation}
\label{eq:eq3}
T_C^{MC*} = \frac{S+1}{S} T_C^{MC}
\end{equation}
Here, $S$ represents the nominal spin magnetic moment for the ionic magnetic atom in each compound. 
We represent this corrected approach in the figure as [GGA+U]*. By applying this correction, the MAPE is further reduced to 44\%, 
significantly improving the accuracy of our transition temperature predictions.
In [GGA+U]*,  $~45\%$ of compounds have absolute percentage error (APE) less than 20$\%$,  
$~25\%$ of compounds have APE between $20\%$ and $40\%$ and  
$~20\%$ compounds have APE between $40\%$ and $80\%$. 
There are three compounds (i.e., Fe$_2$TeO$_6$, Li$_2$MnO$_3$, and MnWO$_4$) 
that have APE larger than $100\%$. The worst case is Li$_2$MnO$_3$ (with 270 APE),  
where GGA+$U$ predicts the compound as ferromagnet while the experiment indicates 
an antiferromagnetic ground state for the compound~\cite{Lee_2012}.
The other ab initio calculation based on DFT+$U$ 
also wrongly predicts this material as a ferromagnet~\cite{Korotin2015}.	
This is because of badly overestimation of Hubbard $U$ for this compound. 
Our investigation finds that the $U$ parameter should be less than 2.01 eV 
for this compound to have an antiferromagnetic ground state. 
In comparison, the self-consistent linear response method overestimates it as 6.39 eV.

On average, according to our data, self-consistent Hubbard $U$ is ~0.5 eV smaller than the non-self-consistent Hubbard $U$. 
Since larger Hubbard $U$ decreases the antiferromagnetic interaction, and in GGA+$U$ for most cases, 
there is an underestimation of antiferromagnetic interactions, 
estimation of Hubbard $U$ through the self-consistent process is vital to predicting the transition temperature.

Although [GGA+U]* produces reasonably accurate results for approximately 50\% of cases, 
it also exhibits significant errors in some instances. 
To better understand the reasons behind these discrepancies, 
we conducted an analysis to investigate whether there exists any meaningful correlation between the APE 
of [GGA+U]* and various compound properties, such as the magnetic moment. 
To do this, we considered several properties, including the Shannon entropy of d-orbital in density of states (DOS), 
the distance between the center of d-orbital and the center of p-orbital in DOS, band gap, 
magnetic moment, ionic bond percentage, total stress, and force. 
Detailed information on these properties for each compound can be found in Tables A1 and A2 in the Appendix \ref{app:A}.
The results of our analysis indicate that there is no meaningful correlation between these properties 
and the APE of the transition temperature derived from [GGA+U]*. 

We also investigate the extended Hubbard model, so-called GGA+$U$+$V$~\cite{DFTUV1,DFTUV2,DFTUV3} 
containing both on-site ($U$) and intersite electronic interactions ($V$), on only 18 compounds. 
If we still consider the $(S+1)/S$ correction for analysis of GGA+$U$+$V$ results, 
there is no critical gain for transition temperature compared to GGA+$U$, on average. 
The results of GGA+$U$ and GGA+$U$+$V$ on the prediction of transition temperatures are collected in Tables A3 and A4 of appendix \ref{app:C}.
\begin{table*}
	\caption{
	The table shows transition temperatures for a subset of samples, with and without geometry optimization. 
	$T_{U}$ and $T_{UV}$ represent the transition temperatures (N\'eel temperatures) obtained 
	using the GGA+$U$ and GGA+$U$+$V$ calculations without geomertry optimization, respectively. 
	The Hubbard parameters $U$ and $V$ (on-site $U$ and inter-site $V$) are obtained from self-consistent linear response theory calculations. 
	On the other hand, $T_{U}^{\mathrm{opt}}$ and $T_{UV}^{\mathrm{opt}}$ indicate N\'eel temperatures obtained from GGA+$U$ and GGA+$U$+$V$ calculations         through geometry optimization along with self-consistent linear response theory calculations for computing $U$ and $V$. 
	The values inside parentheses indicate the temperatures after applying the $\frac{S+1}{S}$ correction factor. 
	The last column presents the experimental temperature value documented in the literature.
	}
	\label{tab:opt}
	\centering
	\begin{tabular}{@{\hspace{3mm}} c @{\hspace{3mm}} c @{\hspace{3mm}} c @{\hspace{3mm}} c @{\hspace{3mm}} c @{\hspace{3mm}} c @{\hspace{3mm}} c @{\hspace{3mm}} c}
		\hline
		&Sample & S &  $T_{U}(K)$  &  $T_{UV}(K)$  & $T_{U}^{\mathrm{opt}}$(K) &  $T_{UV}^{\mathrm{opt}}$(K)  &  $T_{Exp.}(K)$  \\
		\hline
		&NiO      & 1             &  $220\hspace{3mm}(440)$ &  $246\hspace{3mm}(492)$ &  $210\hspace{3mm}(420)$  &  $236\hspace{3mm}(472)$  & $523$\cite{hutchings1972measurement} \\
		&MnO      & $\frac{5}{2}$ & \hspace{4mm}$27\hspace{3mm}(37.8)$  &  \hspace{4mm}$43\hspace{3mm}(60.2)$ &  \hspace{5mm}$72\hspace{3mm}(100.8)$ &  \hspace{5mm}$87\hspace{3mm}(121.8)$ & $117$\cite{kohgi1972inelastic}\\
		&MnS      & $\frac{5}{2}$ &  \hspace{4mm}$54\hspace{3mm}(75.6)$ &  \hspace{4mm}$60\hspace{3mm}(84.0)$ &  \hspace{4mm}$58\hspace{3mm}(81.2)$  & \hspace{4mm}$65\hspace{3mm}(91.0)$   & $152$\cite{clark2021inelastic}\\
		&MnSe     & $\frac{5}{2}$ & \hspace{1mm} $25\hspace{3mm}(35)$   &  \hspace{4mm}$26\hspace{3mm}(36.4)$ &  \hspace{4mm}$37\hspace{3mm}(51.8)$  &  \hspace{4mm}$44\hspace{3mm}(61.6)$  & $124$\cite{milutinovic2002raman}\\
		&NiF$_2$  & 1             &  \hspace{1mm}$37\hspace{3mm}(74)$   &  \hspace{2mm}$44\hspace{3mm}(88)$   &  \hspace{2mm}$32\hspace{3mm}(64)$    & \hspace{2mm}$40.2\hspace{3mm}(80.4)$ & $68.5$\cite{fleury1969paramagnetic}\\
		&MnF$_2$  & $\frac{5}{2}$ &  \hspace{5mm}$57\hspace{3mm}(79.8)$ &  \hspace{5mm}$67\hspace{3mm}(93.8)$ &  \hspace{5mm}$44\hspace{3mm}(61.6)$  &  \hspace{5mm}$57\hspace{3mm}(79.8)$  & $67.3$\cite{nordblad1981specific}\\
		&MnTe     & $\frac{5}{2}$ & \hspace{5mm}$189\hspace{3mm}(264.6)$& \hspace{5mm}$196\hspace{3mm}(274.4)$&  \hspace{5mm}$152\hspace{3mm}(212.8)$&  \hspace{5mm}$156\hspace{3mm}(218.4)$& $310$\cite{szuszkiewicz2006spin}\\
		&CrCl$_2$ & 2             & \hspace{2mm}$2\hspace{3mm}(3)$      &  \hspace{2mm}$2\hspace{3mm}(3)$     &  \hspace{5mm}$19\hspace{3mm}(28.5)$  &  \hspace{5mm}$16\hspace{3mm}(24.0)$  & $11.3-16$\cite{winkelmann1997structural}  \\
		\hline
		&MAPE &              & \hspace{2mm}$58\:\%\hspace{3mm}(41\:\%)$      &  \hspace{3mm}$52\:\%\hspace{3mm}(41\:\%)$     &  \hspace{3mm}$51\:\%\hspace{3mm}(37\:\%)$  &  \hspace{3mm}$41\:\%\hspace{3mm}(31\:\%)$  & -  \\
		\hline
	\end{tabular}
\end{table*}

Ultimately, we explore the effect of geometry optimization of primitive antiferromagnetic cells (atomic positions and crystal lattice vectors) 
to determine if the consistent crystal structures with GGA+$U$ and GGA+$U$+$V$ lead to more corrected results for transition temperature.  
For this purpose, we choose 8 compounds with low computational cost, then optimize their structures and then re-calculate $U$ and $V$ parameters for the optimized structures. We repeat this procedure until we reach to optimized structures with self-consistent $U$ and $V$ parameters.
We collect the results in Table ~\ref{tab:opt} with $(S+1)/S$ correction inside the parenthesis. 
Without geometry optimization and considering $(S+1)/S$ correction, 
GGA+$U$ and GGA+$U$+$V$ have almost similar results. 
Both GGA+$U$ and GGA+$U$+$V$ for these 8 compounds (using self-consistent $U$ and $V$) indicate 
41\% MAPE for predicting transition temperatures, comparable with MAPE obtained for all 29 compounds. 
Using optimized structures decrees MAPE  to  37\% and 31\% for GGA+$U$ and GGA+$U$+$V$ (with $(S+1)/S$ correction), respectively. 
Therefore consistent structures with GGA+$U$ and GGA+$U$+$V$ theories can be beneficial for better prediction.
\section{Conclusions}
The study conducted systematic research to benchmark three different Density Functional Theory (DFT) methods - GGA, GGA+$U$, and GGA+$U$+$V$ - for predicting transition temperatures in a group of antiferromagnetic insulators. The results indicate that obtaining Hubbard $U$ parameters through self-consistent processes using linear response theory is crucial for accurate predictions. Additionally, optimizing the structures of the materials leads to more consistent outcomes with experimental data.
To enhance the accuracy of GGA+$U$ results, the study recommends applying a $(S+1)/S$ correction to the GGA+$U$ approach. Moreover, the work introduces a method to determine the appropriate number of nearest neighbors to calculate exchanges within a supercell.
In conclusion, this study highlights the performance of various DFT methods in predicting transition temperatures for antiferromagnetic insulators. It also proposes correction strategies to improve the accuracy of these predictions, providing valuable insights for future research in this area.
\begin{acknowledgements}
The authors thank Iurii Timrov for helping them with GGA+$U$ and GGA+$U$+$V$ calculations.  
M. A. thanks Majid Gazor for introducing the null space concept in linear algebra and Farhad Shahbazi 
for suggesting the Shannon entropy to measure the localization of d-orbitals in DOS.
\end{acknowledgements}

\setcounter{equation}{0}
\setcounter{figure}{0}
\setcounter{table}{0}
\renewcommand{\theequation}{A\arabic{equation}}
\renewcommand{\thetable}{A\arabic{table}}
\renewcommand{\thefigure}{A\arabic{figure}}

\appendix

\section{Characterization of compounds}
\label{app:A}
\subsection{Hubbard parameter}
\label{app:A1}
The Hubbard $U$ parameters are determined using linear response density-functional perturbation theory (LR-DFPT). 
Typically, the $U$ value is computed with a precision of approximately $0.01$ eV, 
and this calculation is carried out iteratively until convergence is achieved. 
The calculated $U$ parameters for various compounds are summarized in Table ~\ref{tab:type}.
\subsection{Ionic percentage and magnetic moment}
\label{app:A2}
To describe the covalent and ionic nature of bonds, the net charges of ions are calculated using Bader charge analysis \cite{bader1985atoms} using bader code \cite{Code}.   
The results evident that net charge on  atoms in compounds are less than nominal charges.  
For example in NiO  about 1.26 electrons transfer from each Ni atom to O atoms while  nominal charges is $+2$ for Ni and $-2$ for O ions. 
Therefore, in NiO, the bond between Ni and O atom shows obvious ionic character and a little covalent bond (37 percentage).
The ionic percentage of different compounds is shown in Table \ref{tab:type}.
The results reveal that in all compounds, the bonding behavior consists of a combination of ionic and covalent bonds. 
When comparing the results obtained from GGA and GGA+$U$ calculations, 
it becomes evident that the degree of ionic bonding increases across all compounds when using the GGA+$U$ approach. 
This increase in ionic character is a result of correcting the on-site Coulomb interaction, which aligns with the findings in Ref.~\cite{das2021first}.

We also determine the magnetic moment through Bader charge analysis, 
using two different approximations: GGA and GGA+$U$. 
Notably, the magnetic moment exhibits an increase when employing the GGA+$U$ approximation, as depicted in Table \ref{tab:type}. 
This phenomenon can be attributed to the relationship between the magnetic moments of the transition metal atoms and their localized d electron shells. 
The GGA+$U$ approximation increases d electron shells localization and eliminates fractional occupation numbers, 
which is why we anticipate an increase in the magnetic moment when utilizing this approach \cite{tesch2022hubbard}.
\begin{table*}[]
	\caption{Hubbard parameter, magnetic moment, total stress, total force, 
	         ionic percentage and charge distribution for all samples within GGA+$U$ approximation. 
		 The data inside the parentheses represent the calculated magnetic moment and ionic percentage in the GGA approach.}
	\label{tab:type}
	\begin{tabular}{lccccccc}
		\hline
		 Sample& \makecell{Hubbard parameter\\ (eV)} & \makecell{Magnetic moment\\ ($\mu{_B})$} & \makecell{Total stress\\ ($\frac{Ry}{bohr\:^3}$)} & \makecell{Total force\\ ($\frac{Ry}{bohr}$)} & Ionic percentage &  Charge distribution\\
		\hline
		NiO            &  $7.17$   & $1.72$\:($1.33$)&$104.84$  & 0 &$63$\:($56$)&Ni$^{+1.26}$  O$^{-1.25}$\\
		MnO            &  $4.81$   & $4.61$\:($4.40$)&$84.99$  & 0 &$72.5$\:($69.5$)&Mn$^{+1.45}$  O$^{-1.44}$\\
		MnS            &  $4.39$   & $4.61$\:($4.33$) &$30.63$  & 0 &$63$\:($58.5$)&Mn$^{+1.26}$  S$^{-1.25}$\\
		MnSe           &  $4.19$   & $4.62$\:($4.35$)&$30.31$  & 0 &$57.5$\:($52.5$)&Mn$^{+1.15}$  Se$^{-1.14}$\\
		Cr$_2$O$_3$    &  $5.93$   & $2.88$\:($2.63$)    &$152.96$  &$0.033$ &$62.3$\:($58$) &Cr$^{+1.85}$  O$^{-1.23}$\\
		Fe$_2$O$_3$    &  $6.18$   & $4.25$\:($3.40$)    &$219.66$  &$0.316$ &$61.6$\:($52.6$) &Fe$^{+1.86}$  O$^{-1.24}$\\
		BiFeO$_3$      &  $6.13$   & $4.25$\:($3.81$)    &$23.05$  &$0.027$ &$62.3$\:($55$) &Bi$^{+1.86}$ Fe$^{+1.87}$O$^{-1.24}$\\
		NiBr$_2$       &  $6.31$   & $1.66$\:($1.39$)    &$1.34$  &$0.001$ &$46.5$\:($38.5$)&Ni$^{+0.93}$  Br$^{-0.46}$\\
		YVO$_3$        &  $4.74$   & $1.81$\:($1.55$)     &$81.33$  &$0.037$ &$64$\:($60$)&Y$^{+2.16}$ V$^{+1.92}$O$^{-1.36}$\\
		LiMnPO$_4$     &  $4.20$   & $4.68$\:($4.58$)    &$55.56$  &$0.252$ &$78.5$\:($75.5$)&Li$^{+0.9}$ Mn$^{+1.57}$P$^{+4.88}$O$^{-1.7}$\\
		LiNiPO$_4$     &  $7.22$   & $1.80$\:($1.58$)      &$75.45$  &$0.128$ &$69$\:($63$)&Li$^{+0.9}$ Ni$^{+1.38}$P$^{+4.88}$O$^{-1.83}$\\
		LiCoPO$_4$     &  $5.91$   & $2.78$\:($2.60$)      &$54.93$  &$0.053$ &$72$\:($67$)&Li$^{+0.9}$ Co$^{+1.44}$P$^{+4.88}$O$^{-1.84}$\\
		YFeO$_3$       &  $6.22$   & $4.23$\:($3.79$)      &$76.98$  &$0.028$ &$62$\:($55$)&Y$^{+2.17}$ Fe$^{+1.86}$O$^{-1.34}$\\
		LaFeO$_3$      &  $6.26$   & $4.22$\:($3.77$)    &76  &$0.145$ &$61.6$\:($55$)&La$^{+2.09}$ Fe$^{+1.85}$O$^{-1.31}$\\
		LiMnO$_2$      &  $5.97$   & $3.92$\:($3.55$)    &$101.46$  &$0.070$ &$56.6$\:($53$)&Li$^{+0.89}$ Mn$^{+1.70}$O$^{-1.31}$\\
		CrCl$_2$       &  $6.41$   & $3.84$\:($3.69$)    &$41.43$  &$0.018$ &$69.5$\:($63.53$)&Cr$^{+1.39}$ Cl$^{-0.70}$\\
		KNiPO$_4$      &  $7.13$   & $1.81$\:($1.59$)    &$56.18$  &$0.259$ &$67.5$\:($60.5$)&K$^{+0.87}$ Ni$^{+1.35}$P$^{+4.88}$O$^{-1.79}$\\
		MnF$_2$        &  $3.84$   & $4.68$\:($4.59$)    &$56.13$  &$0.002$ &$80.5$\:($78.5$)&Mn$^{+1.60}$  F$^{-0.80}$\\
		NiF$_2$        &  $7.14$   & $1.81$\:($1.62$)      &$55.76$  &$0.006$ &$75$\:($69.5$)&Ni$^{+1.50}$  F$^{-0.74}$\\
		Fe$_2$TeO$_6$  &  $6.12$   & $4.31$\:($3.87$)      &$175.19$  &$0.093$ &$65$\:($58.3$)&Fe$^{+1.95}$ Te$^{+5.92}$ O$^{-1.65}$\\
		La$_2$NiO$_4$  &  $7.58$   & $1.76$\:($0.55$)    &$77.46$  &$0.030$ &$66.5$\:($55.5$)&La$^{+2.01}$ Ni$^{+1.33}$ O$^{-1.34}$\\
		Cr$_2$TeO$_6$  &  $5.89$   & $2.87$\:($2.64$)     &$223.71$   &$0.234$ &$64$\:($59.6$)&Cr$^{+1.92}$ Te$^{+5.9}$ O$^{-1.64}$\\
		KMnSb          &  $4.33$   & $4.70$\:($4.20$)     &$29.83$   &$0.032$ &$34$\:($26.5$) &K$^{+0.72}$ Mn$^{+0.68}$ Sb$^{-1.39}$ \\
		Cr$_2$WO$_6$   &  $6.01$   & $2.89$\:($2.66$)    &$114.6$  &$0.042$ &$65.3$\:($61.6$) &Cr$^{+1.96}$ W$^{+3.10}$ O$^{-1.18}$\\
		NiWO$_4$       &  $7.74$   & $1.84$\:($1.52$)     &$61.22$  &$0.109$&$72.5$\:($65$) &Ni$^{+1.45}$ W$^{+2.99}$ O$^{-1.15}$\\
		MnWO$_4$       &  $5.71$   & $4.61$\:($4.36$)     &$92.63$ &$0.217$ &$81$\:($79.5$) &Mn$^{+1.62}$ W$^{+2.99}$ O$^{-1.16}$\\
		CoWO$_4$       &  $6.53$   & $2.78$\:($2.57$)     &$100.78$   &$0.080$ &$73$\:($69.5$) &Co$^{+1.46}$ W$^{+3.02}$ O$^{-1.16}$\\
		Li$_2$MnO$_3$  &  $6.39$   & $3.27$\:($2.68$)    &$87.54$  &$0.065$ &$47.5$\:($45.2$) &Li$^{+0.88}$ Mn$^{+1.90}$ O$^{-1.24}$\\
		MnTe           &  $4.05$   & $4.63$\:($4.28$)     &$26.42$   &0  &$47.5$\:($41.5$) &Mn$^{+0.95}$  Te$^{-0.96}$\\
		\hline
	\end{tabular}
\end{table*}%

\subsection{Projected density of states and Band gap}
\label{app:A3}
 We present the project density of states (PDOS) for all compounds in Figures  S1 and S2 of
 the Supplemental Material using GGA+$U$ approximation.
To convert PDOS information to numbers, we define  Shannon entropy to measure the localization of  the d-orbital of magnetic atoms in energy space:
\begin{equation}
\label{eq:s}
	S_{\mathrm{PDOS}}^{d} = -\sum_{E<E_{\mathrm{Fermi}}}  \rho_d(E) \ln \rho_d(E)
\end{equation}
Here, $\rho_d(E)$ is the normalized DOS of d-orbitals of magnetic atoms, and $E_{\mathrm{Fermi}}$ is Fermi level energy.
Additionally, we calculate the distance between the center of d-orbitals of magnetic moment 
and the p-orbitals of anions in the energy space (d-p distance) to give a measure of the separation of these orbitals. 
This information is gathered in Table ~\ref{tab:gap}.

\begin{figure}[]
        \includegraphics[scale=0.650]{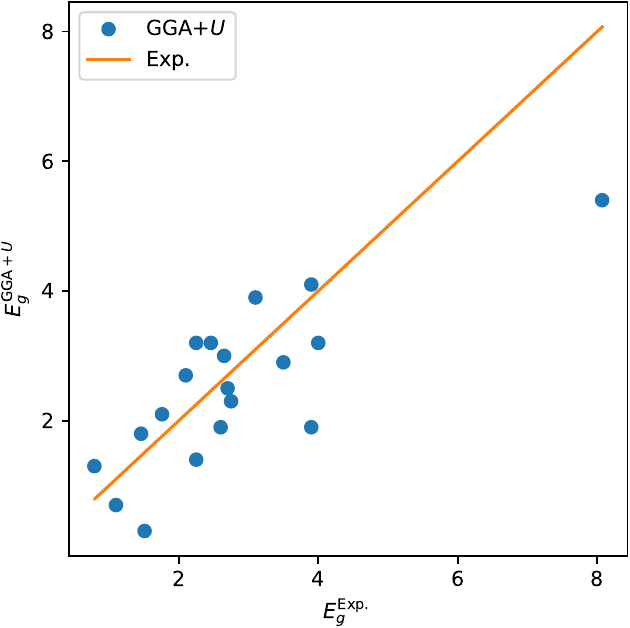}
	\caption{ Comparison of GGA+$U$ band gap ($E_{g}^{\mathrm{GGA}+U}$) with experimental values 
		  ($E_{g}^{\mathrm{Exp.}}$).}
	\label{fig:S3}
\end{figure}

We also report the band gap obtained from the GGA+$U$ calculation in Table ~\ref{tab:gap}. 
We add to the table the existing gap reported in the experiments. 
The gap error using GGA+$U$ for these materials, on average, is about 30\%. 
For some materials, GGA+$U$ overestimates the gap, and for others underestimates the gap in comparison with the experiment (Figure ~\ref{fig:S3}).
\begin{table}[h]
	\centering
	\caption{The Shannon entropy of d-orbitals ($S_{\mathrm{PDOS}}^{d}$) defined in Eq. ~\ref{eq:s} obtained from PDOS, 
		the distance between $p$ orbital center  and $d$ orbital center(d-p distance)  
		, GGA+$U$ band gap  $E_{g}^{GGA+U}$ for all samples and  the reported experimental value of the band gap $E_{g}^{Exp}$.}
	\label{tab:gap}
	\begin{tabular}[t]{lccccc}
		\hline
		Sample& $S_{\mathrm{PDOS}}^{d}$  &\makecell{d-p distance\\ (${eV}$)}& \makecell{$E_{g}^{GGA+U}$\\ (${eV}$)}  & \makecell{$E_{g}^{Exp.}$\\ (${eV}$)} \\
		\hline
		NiO             &\hspace{3mm}$4.05$ &\hspace{3mm}$0.82$  &\hspace{3mm}$3.2$   & \hspace{3mm}$4.0$\cite{bennett2019systematic}\\
		MnO             &\hspace{3mm}$3.96$ &\hspace{3mm}$0.25$ &\hspace{3mm}$1.9$   & \hspace{3mm}$3.9$\cite{bennett2019systematic} \\
		MnS             &\hspace{3mm}$3.44$ &\hspace{3mm}$1.08$‌ &\hspace{3mm}$2.3$   &\hspace{3mm}$2.7- 2.8$\cite{youn2004crossroads} \\
		MnSe            &\hspace{3mm}$3.14$ &\hspace{3mm}$1.39$ &\hspace{3mm}$1.4$   & \hspace{3mm}$2-2.5$\cite{youn2004crossroads}\\
		Cr$_2$O$_3$     &\hspace{3mm}$4.08$ &\hspace{3mm}$0.80$ &\hspace{3mm}$3.9$   & \hspace{3mm}$3.1$\cite{singh2019effect}\\
		Fe$_2$O$_3$     &\hspace{3mm}$3.70$ &\hspace{3mm}$3.26$  &\hspace{3mm}$1.9$  & \hspace{3mm}$2.6$\cite{piccinin2019band}\\
		BiFeO$_3$       &\hspace{3mm}$2.92$ &\hspace{3mm}$3.66$ &\hspace{3mm}$2.7$   & \hspace{3mm}$2.1$\cite{mcdonnell2013photo} \\
		NiBr$_2$        &\hspace{3mm}$3.39$ &\hspace{3mm}$2.85$ &\hspace{3mm}$2.5$   &\hspace{3mm}$-$\\
		YVO$_3$         &\hspace{3mm}$4.02$ &\hspace{3mm}$1.05$  &\hspace{3mm} $2.8$   &\hspace{3mm}$-$\\
		LiMnPO$_4$      &\hspace{3mm}$3.97$ &\hspace{3mm}$0.68$ &\hspace{3mm}$4.1$   & \hspace{3mm}$3.8-4.0$\cite{zhou2004electronic}\\
		LiNiPO$_4$      &\hspace{3mm}$4.05$ &\hspace{3mm}$1.28$  &\hspace{3mm}$4.7$   &\hspace{3mm}$-$ \\
		LiCoP$O_4$      &\hspace{3mm}$4.18$ &\hspace{3mm}$0.38$  &\hspace{3mm}$4.5$   &\hspace{3mm}$-$\\
		YFe$O_3$        &\hspace{3mm}$3.08$ &\hspace{3mm}$2.96$  &\hspace{3mm}$3.2$   & \hspace{3mm}$2.46$\cite{wang2019structural}\\
		LaFe$O_3$       &\hspace{3mm}$3.08$ &\hspace{3mm}$0.36$ &\hspace{3mm}$3.0$   & \hspace{3mm}$2.65$\cite{boateng2017dft+}\\
		LiMn$O_2$       &\hspace{3mm}$4.13$ &\hspace{3mm}$1.42$ &\hspace{3mm}$1.3$   & \hspace{3mm}$0.79$\cite{huang2008competition}\\
		CrCl$_2$        &\hspace{3mm}$3.70$ &\hspace{3mm}$0.56$ &\hspace{3mm}$3.8$   &\hspace{3mm}$-$\\
		KNiP$O_4$       &\hspace{3mm}$3.92$ &\hspace{3mm}$0.14$ &\hspace{3mm}$3.8$   &\hspace{3mm}$-$\\
		MnF$_2$         &\hspace{3mm}$3.61$ &\hspace{3mm}$1.36$ &\hspace{3mm}$3.6$&\hspace{3mm}$-$\\
		NiF$_2$         &\hspace{3mm}$4.01$ &\hspace{3mm}$1.00$   &\hspace{3mm}$5.4$   & \hspace{3mm}$8.07$\cite{arumugam2019complex}\\
		Fe$_2$TeO$_6$   &\hspace{3mm}$3.50$ &\hspace{3mm}$3.32$  &\hspace{3mm}$1.8$   & \hspace{3mm}$1.46$\cite{behera2016structural}\\
		La$_2$NiO$_4$   &\hspace{3mm}$4.22$ &\hspace{3mm}$2.06$  &\hspace{3mm}$0.3$   & \hspace{3mm}$1.51$\cite{laouici2021elaboration}\\
		Cr$_2$TeO$_6$   &\hspace{3mm}$4.17$ &\hspace{3mm}$0.18$   &\hspace{3mm}$2.5$   &\hspace{3mm}$-$\\
		KMnSb           &\hspace{3mm}$2.29$ &\hspace{3mm}$3.35$  &\hspace{3mm}$1.2$   &\hspace{3mm}$-$\\
		Cr$_2$WO$_6$    &\hspace{3mm}$4.21$ &\hspace{3mm}$1.45$  &\hspace{3mm}$2.0$   &\hspace{3mm}$-$\\
		NiWO$_4$        &\hspace{3mm}$4.13$ &\hspace{3mm}$1.73$  &\hspace{3mm}$2.9$   & \hspace{3mm}$3.4-3.6$\cite{shepard2018ab}\\
		MnWO$_4$        &\hspace{3mm}$4.17$ &\hspace{3mm}$1.00$ &\hspace{3mm}$2.5$   & \hspace{3mm}$2.7$\cite{peixoto2022experimental}\\
		CoWO$_4$        &\hspace{3mm}$4.20$ &\hspace{3mm}$1.34$  &\hspace{3mm}$3.2$   & \hspace{3mm}$2.25$\cite{bandiello2021electronic}\\
		Li$_2$MnO$_3$   &\hspace{3mm}$3.33$ &\hspace{3mm}$2.40$  &\hspace{3mm}$2.1$   & \hspace{3mm}$1.76$\cite{ranjeh2020edta}\\
		MnTe            &\hspace{3mm}$2.75$ &\hspace{3mm}$1.84$  &\hspace{3mm}$0.7$   & \hspace{3mm}$0.9-1.3$\cite{youn2004crossroads}\\
		\hline
		MAPE           &\hspace{3mm}$-$ &\hspace{3mm}$-$&\hspace{3mm}$30\:\%$   & \hspace{3mm}$-$\\
		\hline
	\end{tabular}
\end{table}%

\section{Null space analysis}
\label{app:B}
\begin{figure*}
	\centering
	\includegraphics[scale=0.50]{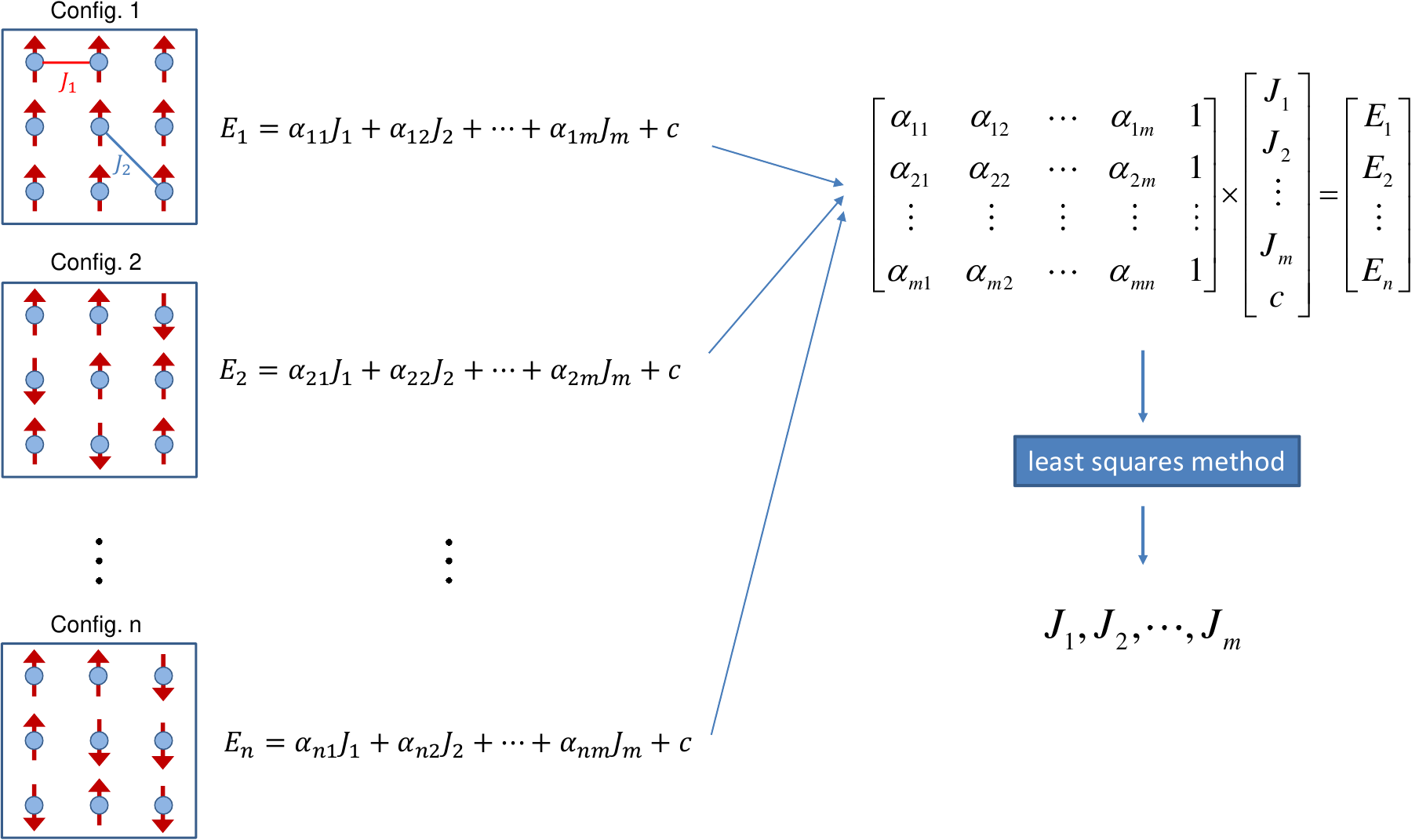}
	\caption {(Color online) The sketch illustrates the least squares method to obtain Heisenberg exchanges. 
		For each colinear magnetic configuration, the total energy of DFT consists of the Heisenberg exchange interactions 
		between magnetic moments plus a constant due to other interactions. So using different configurations and the least squares method, 
		we can obtain exchange parameters.}
	
	\label{fig:lstsq}
\end{figure*}

Fig.~\ref{fig:lstsq} illustrates the calculation of Heisenberg exchange parameters using the DFT total energy of different configurations within a supercell.
In this method, for each magnetic configuration, the Heisenberg Hamiltonian converts to Eq.~\ref{eq:eq2}.
Using supercells to estimate Heisenberg exchange parameters has a bottleneck due to periodic boundary conditions. 
The problem is that when we use a supercell, we can not calculate the exchange parameter for an aribritry distance. 
It is evident that we should restrict exchange parameter estimations within the supercell. 
If $d_{ij}$ is the distance between atom i and atom j, then the supercell method restricts us to estimate only $J_{ij}$ exchange parameters 
that $d_{ij} < L$, where $L$ is supercell size.   
In the first principles ab initio research, people consider this limitation. 
However, the problem is more challenging than it seems. In the following, we will explain it.

According to Eq.~\ref{eq:eq2}, for each magnetic configuration, such as configuration $k$, 
we have a set of coefficients represented as $\alpha_{ki}$.
Using these coefficients, we build a coefficient matrix ($\mathbf{A}_{ki}=\alpha_{ki}$) 
to check the dependency of these matrix columns. 
If column $i$ depends on others, we should restrict our method to calculate the exchange parameter up to $(i-1)$th nearest neighbor. 
In linear algebra, by considering kernel (or null) space, the dependency of columns of a matrix releases. 
The following equation defines the null space of matrix $\mathbf{A}$:
\begin{equation}
\mathrm{Ker}(\mathbf{A})= \{ \mathbf{x}: \mathbf{Ax=0} \}=\mathrm{Null}(\mathbf{A})
\end{equation}
The null space contains solutions for equation $\mathbf{Ax=0}$. 
For a matrix with no-zero $\mathbf{x}$ solution, it means matrix columns are not independent.
To shed light on the issue, we explain it using the NiO example. 
For a $2\times2\times2$ supercell from NiO primitive cell (i.e., a trigonal), 
the matrix of coefficients for 12 (random) different configurations is:
\begin{equation}
\begin{bmatrix}
$96$&  $48$& $192$& $96$&  $192$& $64$& $384$&  $48$&  $288$\\ 
$-16$& $0$&  $0$&   $0$&   $32$&  $0$&  $0$&    $-48$& $16$\\ 
$-32$& $48$& $-64$& $96$&  $-64$& $64$& $-128$& $48$&  $-96$\\ 
$16$&  $0$&  $48$&  $48$&  $64$&  $0$&  $96$&   $0$&   $80$\\ 
$8$&   $12$& $-8$&  $48$&  $-32$& $16$& $-16$&  $0$&   $-24$\\ 
$0$&   $12$& $-24$& $60$&  $-48$& $16$& $-48$&  $12$&  $-48$\\ 
$-8$&  $12$& $-8$&  $48$&  $0$&   $16$& $-16$&  $0$&   $-8$\\ 
$-8$&  $0$&  $8$&   $24$&  $32$&  $0$&  $16$&   $-24$& $24$\\ 
$8$&   $24$& $24$&  $72$&  $32$&  $32$& $48$&   $24$&  $40$\\ 
$24$&  $36$& $48$&  $84$&  $48$&  $48$& $96$&   $36$&  $72$\\ 
$28$&  $24$& $64$&  $60$&  $72$&  $32$& $128$&  $12$&  $100$\\ 
$48$&  $36$& $96$&  $72$&  $96$&  $48$& $192$&  $24$&  $144$\\ 
\end{bmatrix}
\end{equation}
The null space of the marix $\mathbf{A}$ is as follows (by using nullspcae method in scipy.linalg library of Python):
\begin{equation*}
\mathbf{x}=
\begin{bmatrix}
$2$ \\
0 \\
$-2$ \\
0 \\
1 \\
0 \\
0 \\
0 \\
0 
\end{bmatrix}
,
\begin{bmatrix}
0 \\
$-4/3$ \\
0 \\
0 \\
0 \\
1 \\
0 \\
0 \\
0 \\
\end{bmatrix}
,
\begin{bmatrix}
0 \\
0 \\
$-2$ \\
0 \\
0 \\
0 \\
1 \\
0 \\
0 \\
\end{bmatrix}
,
\begin{bmatrix}
$1$ \\
0 \\
$-2$ \\
0 \\
0 \\
0 \\
0 \\
0 \\
1 
\end{bmatrix}
\end{equation*}
The first null space solution, for example, tells us that coefficients in column $5$ depend on columns $1$ and $3$ by the following equation:
\begin{equation*}
2 \times
\begin{bmatrix}
$96$  \\ 
$-16$ \\
$-32$ \\
$16$  \\
$8$   \\
$0$   \\
$-8$  \\
$-8$  \\
$8$   \\
$24$  \\
$28$ \\
$48$ 
\end{bmatrix}
-2 \times
\begin{bmatrix}
$192$  \\
$0$    \\
$-64$  \\
$48$   \\
$-8$   \\
$-24$  \\
$-8$   \\
$8$    \\
$24$   \\
$48$   \\
$64$   \\
$96$ 
\end{bmatrix}
+1\times
\begin{bmatrix}
$192$  \\ 
$32$   \\
$-64$  \\
$64$   \\
$-32$  \\
$-48$  \\
$0$    \\
$32$   \\
$32$   \\
$48$   \\
$72$   \\
$96$  
\end{bmatrix}
=
\begin{bmatrix}
$0$  \\ 
$0$   \\
$0$  \\
$0$   \\
$0$  \\
$0$  \\
$0$    \\
$0$   \\
$0$   \\
$0$   \\
$0$   \\
$0$  
\end{bmatrix}
\end{equation*}
For $2\times2\times2$ supercell, the supercell size apparently allows us to calculate $J_{i}$ until the $12$th nearest neighbor 
(using $d_{ij} < L$ criterion), however, null space analysis warring us to restrict the calculation up to the $4$th nearest neighbor. 
We should increase the supercell size to obtain exchange parameters beyond $4$th nearest neighbor. 
For example, if we use $3\times3\times3$ supercell for NiO, the matrix coefficient for 12 (random) different configurations is as follows:
\begin{equation}
\begin{bmatrix}
$96$&  $48$& $192$& $96$&  $192$& $64$& $384$&  $48$&  $288$\\ 
$-16$& $0$&  $0$&   $0$&   $32$&  $0$&  $0$&    $-48$& $16$\\ 
$-32$& $48$& $-64$& $96$&  $-64$& $64$& $-128$& $48$&  $-96$\\ 
$16$&  $0$&  $48$&  $48$&  $64$&  $0$&  $96$&   $0$&   $80$\\ 
$8$&   $12$& $-8$&  $48$&  $-32$& $16$& $-16$&  $0$&   $-24$\\ 
$0$&   $12$& $-24$& $60$&  $-48$& $16$& $-48$&  $12$&  $-48$\\ 
$-8$&  $12$& $-8$&  $48$&  $0$&   $16$& $-16$&  $0$&   $-8$\\ 
$-8$&  $0$&  $8$&   $24$&  $32$&  $0$&  $16$&   $-24$& $24$\\ 
$8$&   $24$& $24$&  $72$&  $32$&  $32$& $48$&   $24$&  $40$\\ 
$24$&  $36$& $48$&  $84$&  $48$&  $48$& $96$&   $36$&  $72$\\ 
$28$&  $24$& $64$&  $60$&  $72$&  $32$& $128$&  $12$&  $100$\\ 
$48$&  $36$& $96$&  $72$&  $96$&  $48$& $192$&  $24$&  $144$\\ 
\end{bmatrix}
\end{equation}
The null space of the matrix is:
\begin{equation*}
\mathbf{x}=
\begin{bmatrix}
0 \\
1 \\
0 \\
0 \\
$-1/2$ \\
0 \\
0 \\
1 \\
0 
\end{bmatrix}
\end{equation*}
Thus, we can obtain exchange parameters up to the 8th nearest neighbor (since column eight depends on column two and column five).
\newpage
\section{Transition temperature}
\label{app:C}
To determine the transition temperature ($T_C$) for all compounds, we employ classical Monte Carlo (MC) simulations. 
We provide transition temperatures for all compounds along with Heisenberg exchange parameters for GGA and GGA+$U$ approximations 
in Tables S5, S6, S7, and S8 of the Supplemental Material.  

In Table~\ref{tab:temp1}, you can find the transition temperature estimates ($T_C^{MC}$) 
for all 29 compounds obtained through MC simulations using the GGA+U results. 
Additionally, we include the $T_C^{MC}$ correction ($T_C^{MC*}$) by multiplying it by the factor $(S+1)/S$.
Furthermore, in Table \ref{tab:temp2}, we present the transition temperatures for 18 compounds using the GGA+$U$+$V$ approximation. 
It is important to note that the results obtained with the GGA+$U$+$V$ 
approach are comparable to those of GGA+$U$ and do not lead to an improved estimation of the transition temperature.
\begin{table}[h]
	\caption{The transition temperature for all samples was obtained using the Heisenberg exchange interactions calculated from the GGA+$U$ method. $T_{MC}$, was determined through Monte Carlo (MC) simulations. The transition temperature, denoted as $T_{MC}^*$, was then obtained by multiplying the Monte Carlo results with the $\frac{(S+1)}{S}$ factor.  Additionally, $T_{Exp}$ represents the experimentally reported temperature.}
	\label{tab:temp1}
	\centering
	\begin{tabular}{@{\hspace{5mm}} c @{\hspace{5mm}} c @{\hspace{5mm}} c @{\hspace{5mm}} c @{\hspace{5mm}} c @{\hspace{5mm}} c }
		\hline
		Sample & S & \makecell{$T_{MC}$\\ (K)} & \makecell{$T_{MC}^*$\\ (K)} & \makecell{$T_{Exp}$\\ (K)}     \\
		\hline
		NiO       &             1  &  $220$   &  $440$     &  523\cite{hutchings1972measurement}         \\
		MnO       & $\frac{5}{2}$  &  $26$    &  $36.4$    &  117\cite{kohgi1972inelastic}        \\
		MnS       &  $\frac{5}{2}$ &  $54$    &  $75.6$   &  152\cite{clark2021inelastic}        \\
		MnSe      &  $\frac{5}{2}$ &  $25$    &  $35$    &  124\cite{milutinovic2002raman}        \\
		Cr$_2$O$_3$   &  $\frac{3}{2}$ &  $118$   &  $195.8$   &  308\cite{samuelsen1970inelastic}      \\
		Fe$_2$O$_3$   &  $\frac{5}{2}$ &  $701$   &  $981.4$   &  960\cite{samuelsen1970inelastic}       \\
		BiFeO$_3$   &  $\frac{5}{2}$ &  $466$     &  $652.4$  &  650\cite{park2011magnetoelectric}         \\
		NiBr$_2$    &             1  &  $19$    &  $38$    &  52\cite{day1976optical}          \\
		YVO$_3$   &  $\frac{3}{2}$ &     $49.2$      &  $81.6$   &  77\cite{wang2006oxidation}    \\
		LiMnPO$_4$  &  $\frac{5}{2}$ &  $25$   &  $35$   &  34-36\cite{gnewuch2020distinguishing}     \\
		LiNiPO$_4$  &             1  &  $11$   &  $22$   &  $21.8$\cite{kharchenko2003weak}     \\
		LiCoPO$_4$  &  $\frac{3}{2}$ &  $14$   &  $23.2$   &  22-24\cite{gnewuch2020distinguishing}    \\
		YFeO$_3$    &  $\frac{5}{2}$ &  $469$   &  $656.6$   &  $644.5$\cite{shen2009magnetic}    \\
		LaFeO$_3$   &  $\frac{5}{2}$ &  $527$     &  $737.8$   &  750\cite{koehler1957neutron}        \\
		LiMnO$_2$   &             2  &  $111$   &  $166.5$   &  300\cite{kellerman2007structural}      \\
		CrCl$_2$    &             2  &  $2$    &  $3$    &  $11.3-16$\cite{winkelmann1997structural}  \\
		KNiPO$_4$   &             1  &  $9$    &  $18$   &  25\cite{lujan1994magnetic}        \\
		MnF$_2$     &  $\frac{5}{2}$ &  $57$    &  $79.8$    &  $67.3$\cite{nordblad1981specific}     \\
		NiF$_2$     &             1  &  $37$    &  $74$    &  $68.5$\cite{fleury1969paramagnetic}     \\
		Fe$_2$TeO$_6$  &  $\frac{5}{2}$ & $351$    &  $491.4$   &  210\cite{kaushik2016structural}     \\
		La$_2$NiO$_4$ &             1  &  $209$   &  $418$   &  650\cite{lander1989structural}        \\
		Cr$_2$TeO$_6$ &  $\frac{3}{2}$ & $17$    &  $28.2$    &  93\cite{zhu2019amplitude}         \\
		KMnSb     &  $\frac{5}{2}$ &  $217$   &  $303.8$   &  237\cite{mao2020antiferromagnetic}       \\
		Cr$_2$WO$_6$   &  $\frac{3}{2}$ &  $28$   &  $46.4$   &  45\cite{zhu2019amplitude}        \\
		NiWO$_4$    &             1  &  $21$   &  $42$   &  62\cite{prosnikov2017lattice}        \\
		MnWO$_4$    &  $\frac{5}{2}$ &  $20$    &  $28$    &  $8-13.5$\cite{lautenschlager1993magnetic}\\
		CoWO$_4$    &  $\frac{3}{2}$ &  $17$   &  $28.2$    &  40\cite{deng2012preparation}         \\
		Li$_2$MnO$_3$ &  $\frac{3}{2}$ &  $81$   &  $134.4$   &  $36.5$\cite{zhang2011synthesis}    \\
		MnTe      &        $\frac{5}{2}$  & $189$    &  $264.6$  &  310\cite{szuszkiewicz2006spin}        \\
		\hline
		  MAPE    &          &  $53\:\%$  & $44\:\%$   &   -       \\
		\hline
	\end{tabular}
\end{table}
\begin{table}[h]
	\caption{The transition temperature for 18 samples was obtained using the Heisenberg exchange interactions calculated from the GGA+$U$+$V$ method. $T_{MC}$, was determined through Monte Carlo (MC) simulations. The transition temperature, denoted as $T_{MC}^*$, was then obtained by multiplying the Monte Carlo results with the $\frac{(S+1)}{S}$ factor.  Additionally, $T_{Exp}$ represents the experimentally reported temperature.}
	\label{tab:temp2}
	\centering
	\begin{tabular}{@{\hspace{5mm}} c @{\hspace{5mm}} c @{\hspace{5mm}} c @{\hspace{5mm}} c @{\hspace{5mm}} c @{\hspace{5mm}} c }
		\hline
		Sample & S & \makecell{$T_{MC}$\\ (K)} & \makecell{$T_{MC}^*$\\ (K)} & \makecell{$T_{Exp}$\\ (K)}     \\
		\hline
		NiO         &1            & $246$ & $492$    &  523\cite{hutchings1972measurement}         \\
		MnO         &$\frac{5}{2}$& $43$  & $60.2$   &  117\cite{kohgi1972inelastic}        \\
		MnS         &$\frac{5}{2}$& $60$  & $84$     &  152\cite{clark2021inelastic}        \\
		MnSe        &$\frac{5}{2}$& $26$  & $36.4$   &  124\cite{milutinovic2002raman}        \\
		Cr$_2$O$_3$ &$\frac{3}{2}$& $119$ & $178.5$  &  308\cite{samuelsen1970inelastic}      \\
		Fe$_2$O$_3$   &$\frac{5}{2}$& $804$ & $1125.6$ &  960\cite{samuelsen1970inelastic}       \\
		BiFeO$_3$   &$\frac{5}{2}$& $496$ & $694.4$  &  650\cite{park2011magnetoelectric}         \\
		LiMnPO$_4$  &$\frac{5}{2}$& $28$  & $39.2$   &  34-36\cite{gnewuch2020distinguishing}     \\
		YFeO$_3$    &$\frac{5}{2}$& $518$ & $725.2$  &  $644.5$\cite{shen2009magnetic}    \\
		LaFeO$_3$  &$\frac{5}{2}$& $553$ & $774.2$  &  750\cite{koehler1957neutron}        \\
		LiMnO$_2$  &2            & $129$ & $193.5$  &  300\cite{kellerman2007structural}      \\
		CrCl$_2$   &2            & $2$   & $3$      &  $11.3-16$\cite{winkelmann1997structural}  \\
		MnF$_2$    &$\frac{5}{2}$& $72$  & $100.8$  &  $67.3$\cite{nordblad1981specific}     \\
		NiF$_2$    &1            & $47$  & $94$     &  $68.5$\cite{fleury1969paramagnetic}          \\
		NiWO$_4$   &1            & $24$  & $84$     &  62\cite{prosnikov2017lattice}        \\
		MnWO$_4$   &$\frac{5}{2}$& $28$  & $39.2$   &  $8-13.5$\cite{lautenschlager1993magnetic}\\
		CoWO$_4$   &$\frac{3}{2}$& $20$  & $32$     &  40\cite{deng2012preparation}         \\
		MnTe     &$\frac{5}{2}$& $196$ & $274.4$  &  310\cite{szuszkiewicz2006spin}        \\
		\hline
		  MAPE   &          &  $50\:\%$  & $43\:\%$   &   -       \\
		\hline
	\end{tabular}
\end{table} 

\clearpage 
\bibliography{Refs}
\pagebreak
\onecolumngrid
\begin{center}
	\textbf{\large Supplementary information for:Benchmarking density functional theory on the prediction of antiferromagnetic transition temperatures}\\[.2cm]
	Zahra Mosleh,$^{*}$ Mojtaba Alaei,$^{\dag}$ \\[.1cm]
	{\itshape Department of Physics, Isfahan University of Technology, Isfahan 84156-83111, Iran.}		
\end{center}

\setcounter{equation}{0}
\setcounter{figure}{0}
\setcounter{table}{0}
\setcounter{page}{1}
\renewcommand{\theequation}{S\arabic{equation}}
\renewcommand{\thetable}{S\arabic{table}}
\renewcommand{\thefigure}{S\arabic{figure}}

\maketitle

This Supplemental Material presents some computational details of the main manuscript. 
We first provide  projected density of states (PDOS)  of each compound. 
In the following,  we present Hesinebreg exchange parameters derived from GGA using full potential (FPLO code) and pseudopotential method (Quantum-Espresso code). 
Also, we provide exchange parameters obtained from GGA and GGA+$U$ for each compound using Quantum-Espresso. 
\subsection{Projected density of states and Band gap}
\begin{figure}
	\centering
	\subfigure{\includegraphics[width=5cm]{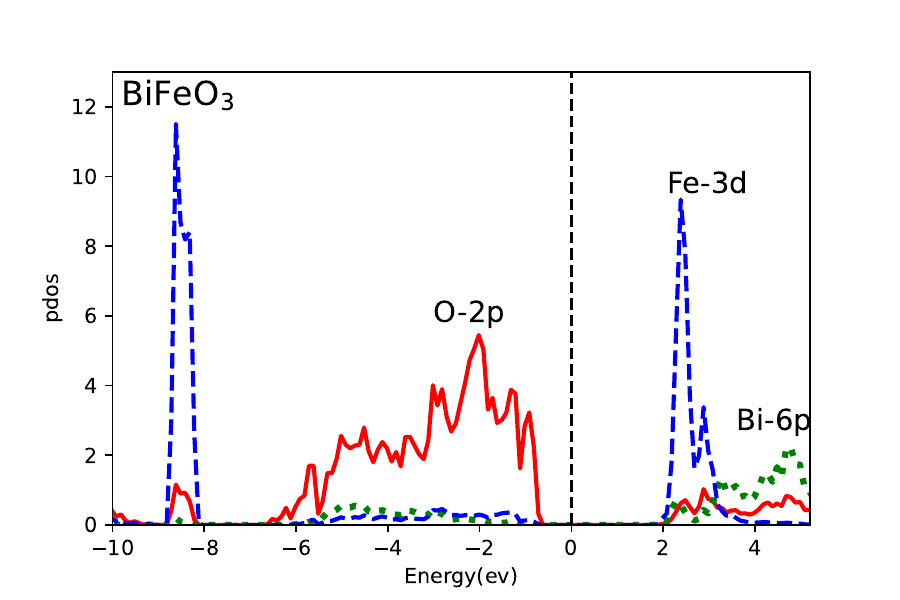}}
	\subfigure{\includegraphics[width=5cm]{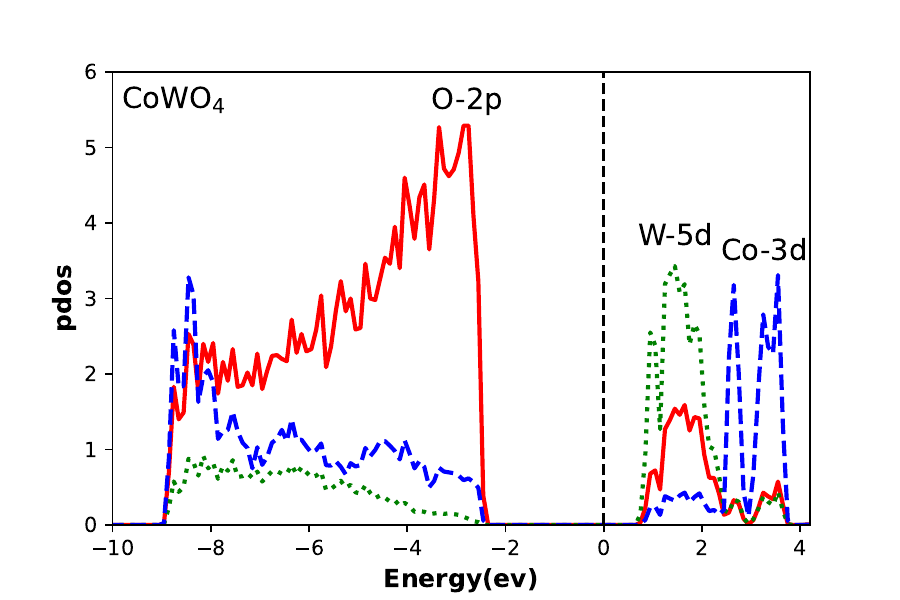}}
	\subfigure{\includegraphics[width=5cm]{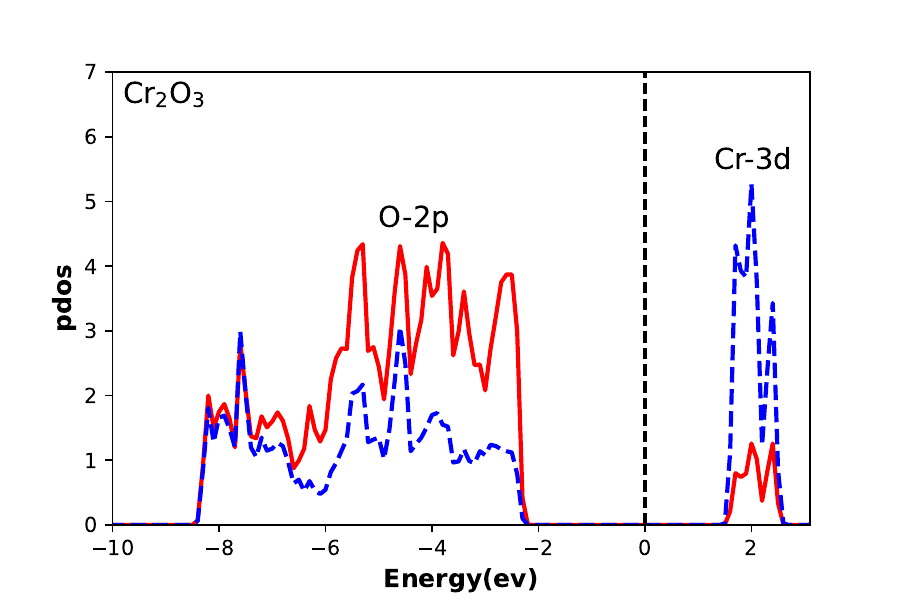}}
	\qquad
	\subfigure{\includegraphics[width=5cm]{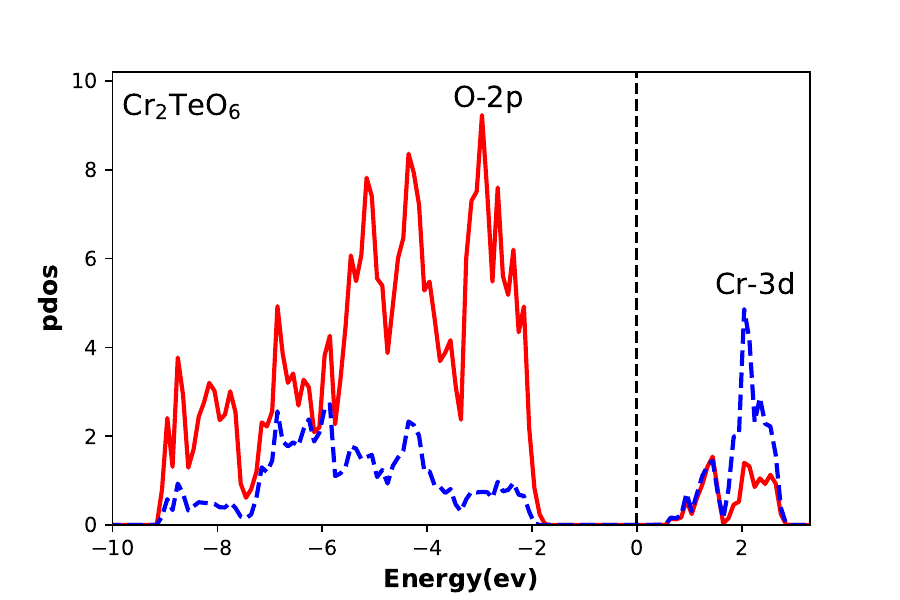}}
	\subfigure{\includegraphics[width=5cm]{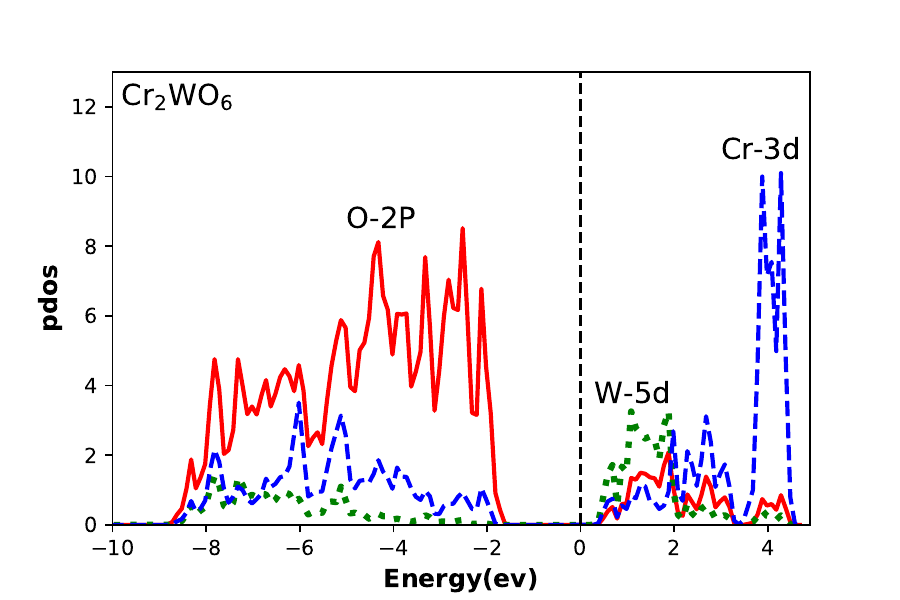}}
	\subfigure{\includegraphics[width=5cm]{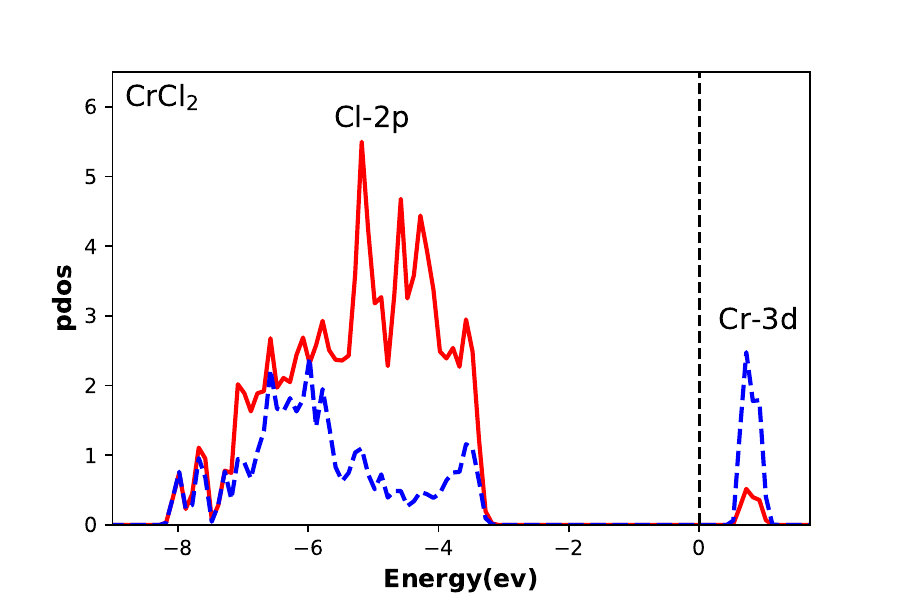}}
	\qquad
	\subfigure{\includegraphics[width=5cm]{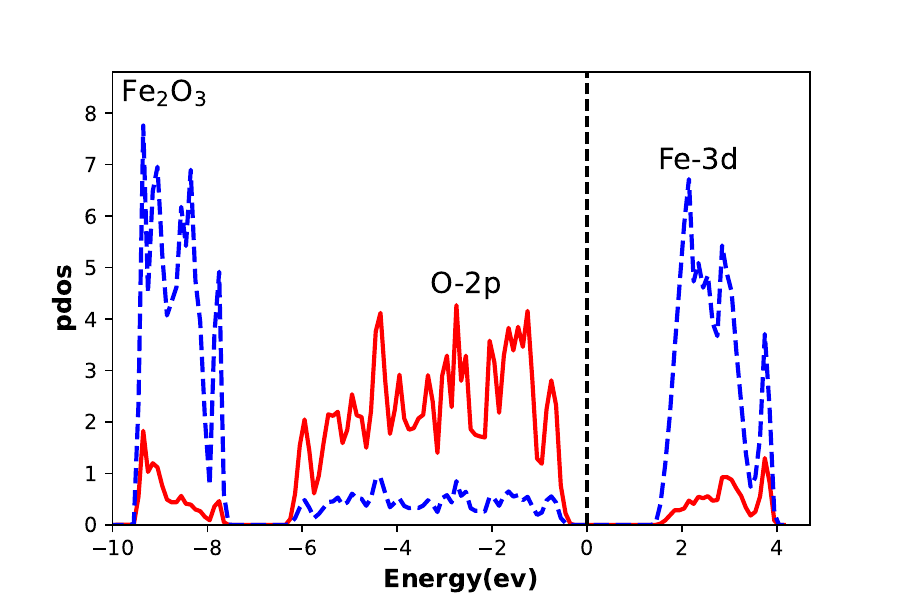}}
	\subfigure{\includegraphics[width=5cm]{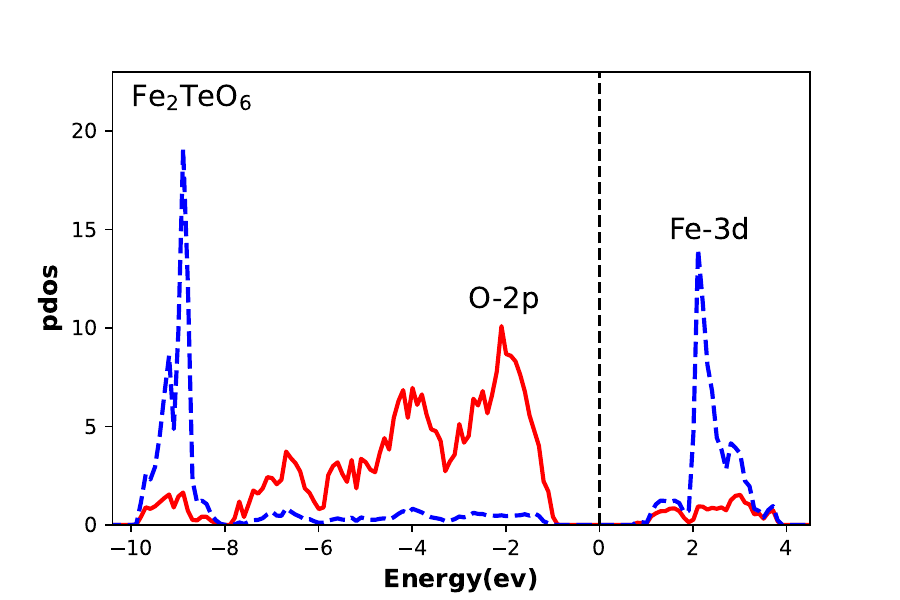}}
	\subfigure{\includegraphics[width=5cm]{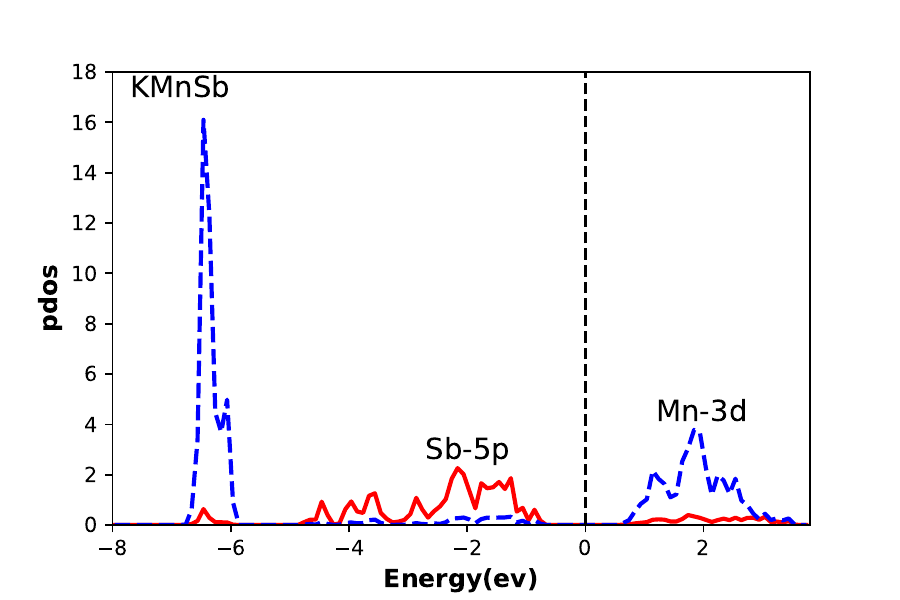}}
	\qquad
	\subfigure{\includegraphics[width=5cm]{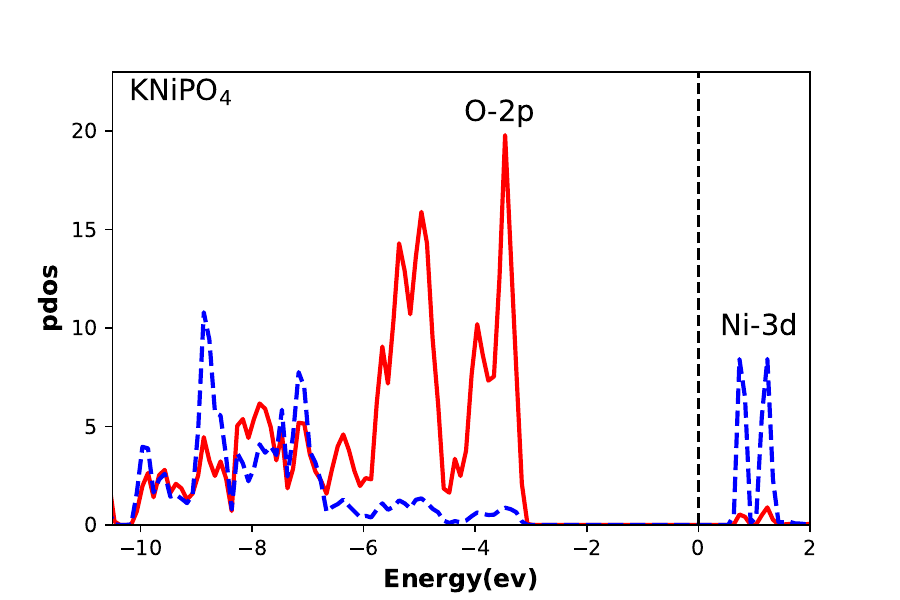}}
	\subfigure{\includegraphics[width=5cm]{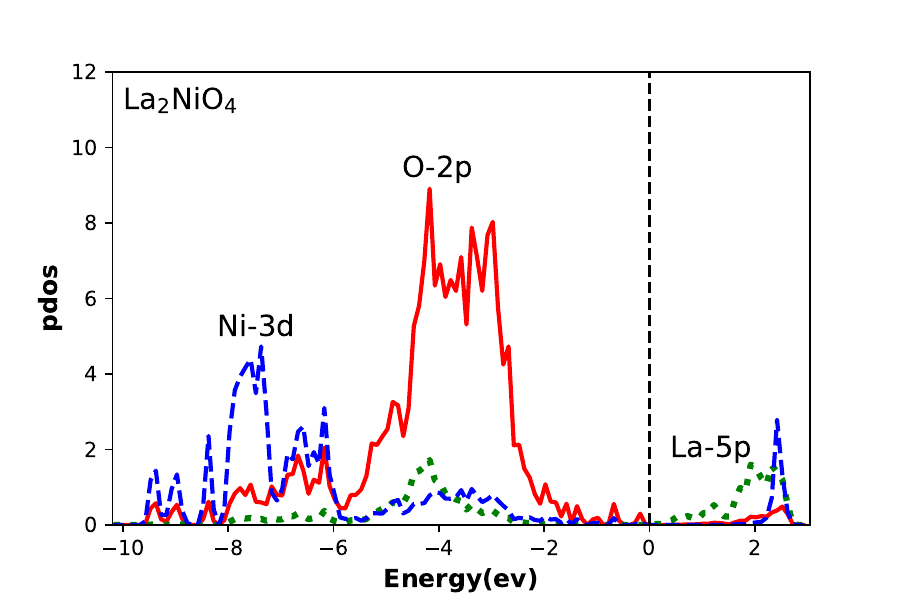}}
	\subfigure{\includegraphics[width=5cm]{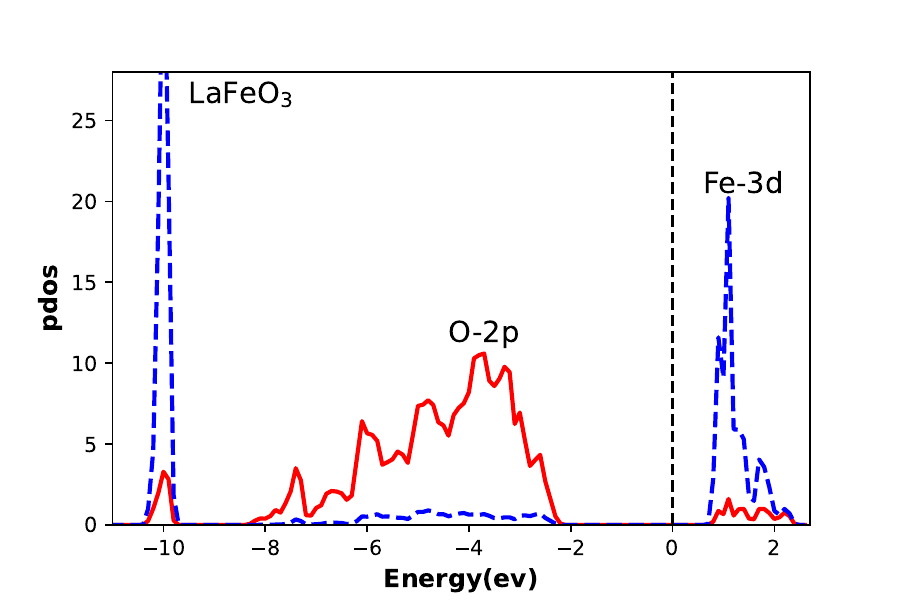}}
	\qquad
	\subfigure{\includegraphics[width=5cm]{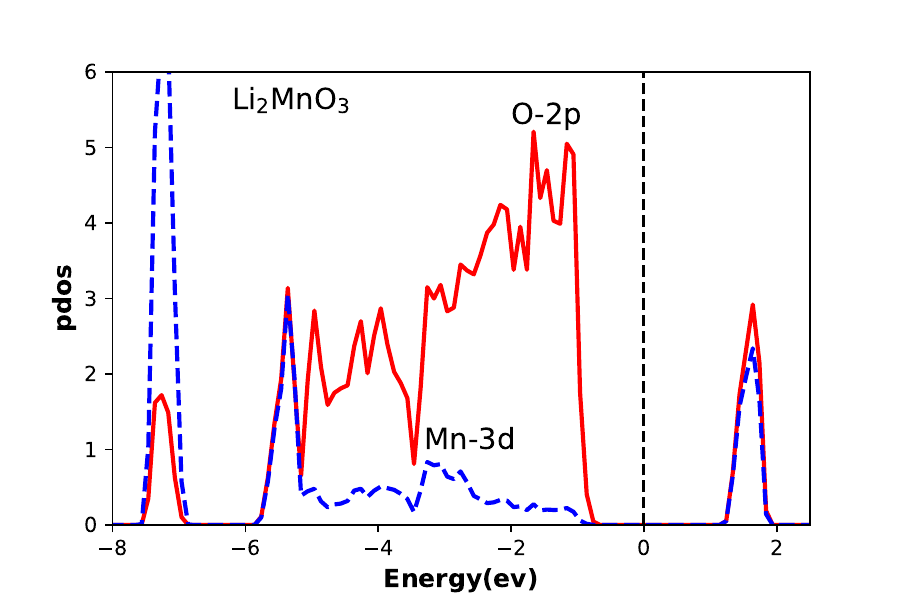}}
	\subfigure{\includegraphics[width=5cm]{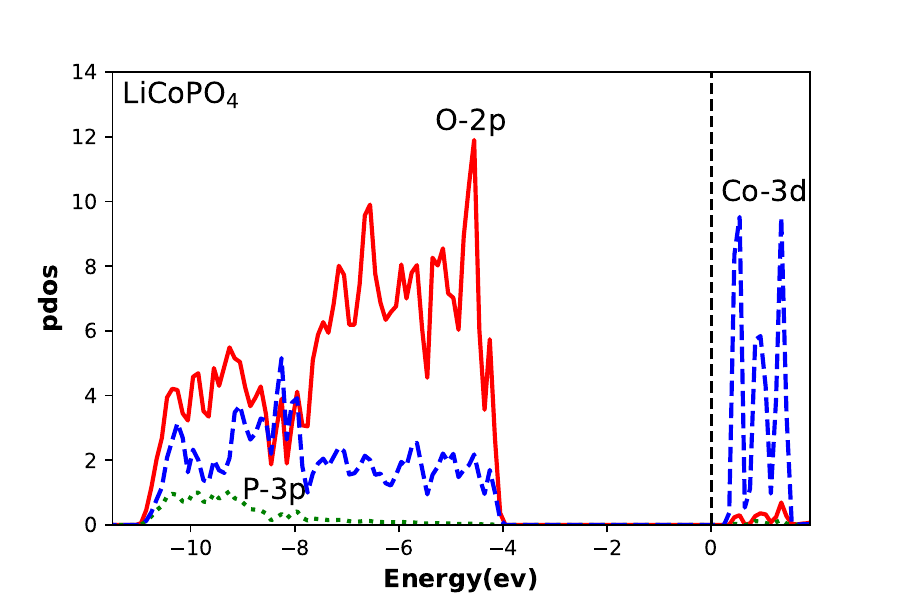}}
	\subfigure{\includegraphics[width=5cm]{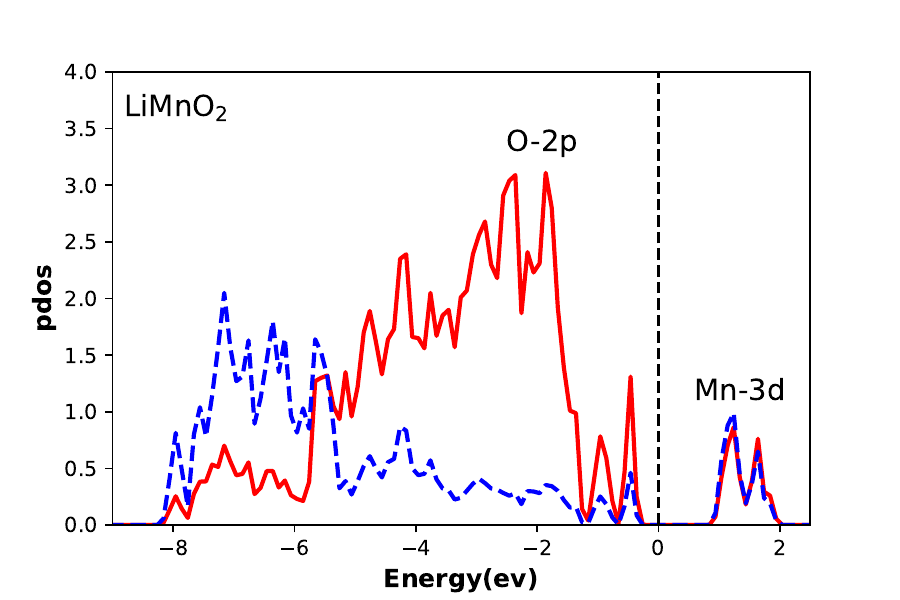}}
	\qquad
	\subfigure{\includegraphics[width=5cm]{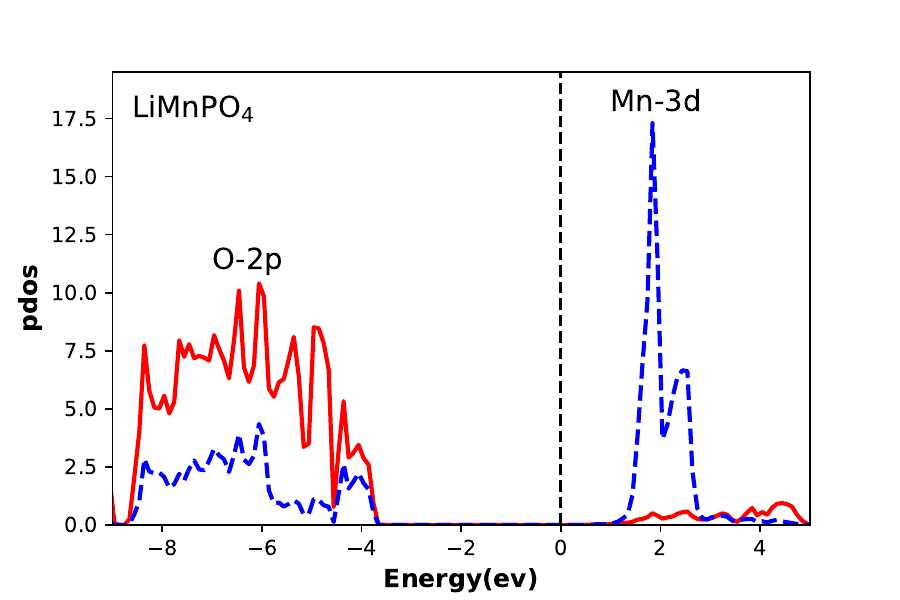}}
	\subfigure{\includegraphics[width=5cm]{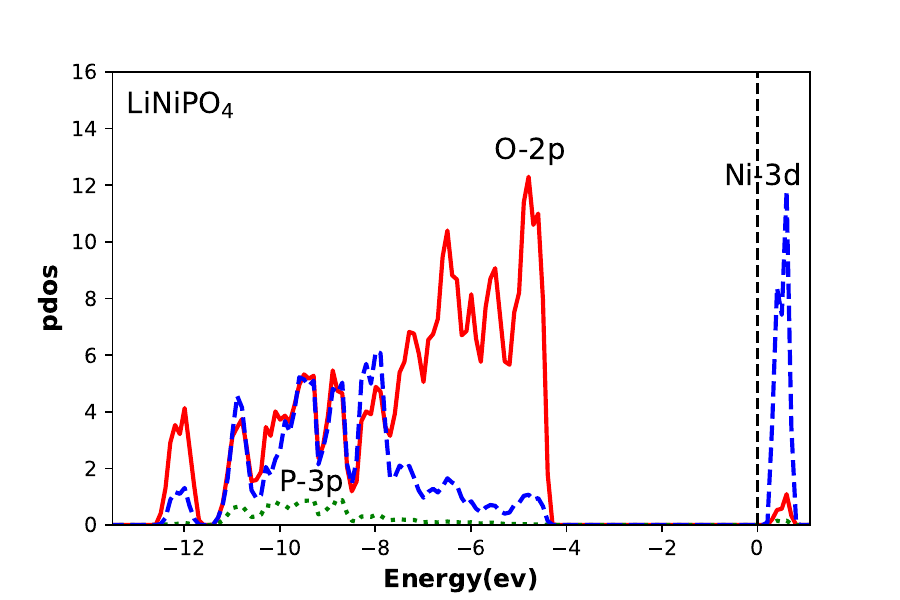}}
	\subfigure{\includegraphics[width=5cm]{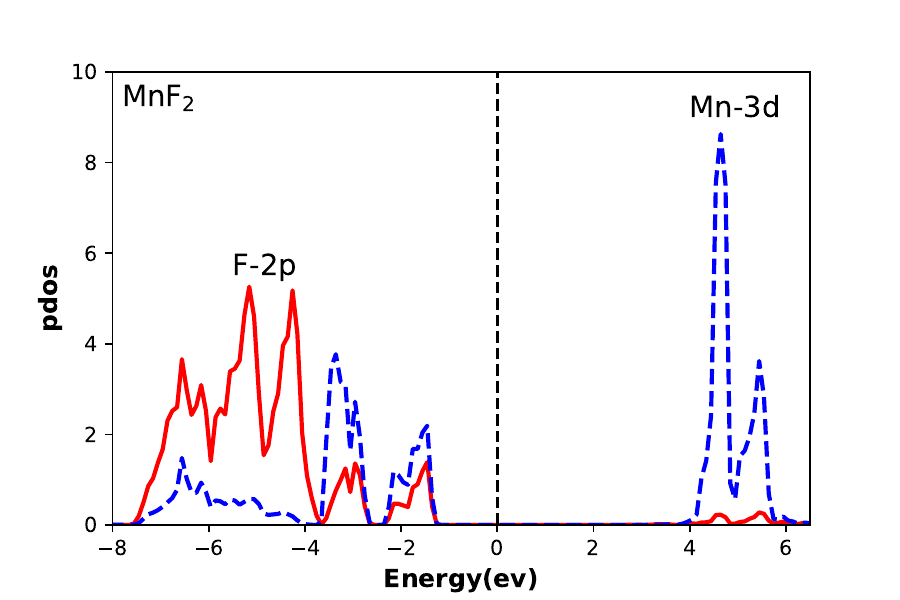}}
	\caption{(Color online) Projected density of states (PDOS) in  GGA+$U$ approximation.}
	\label{fig:S1}
\end{figure}

\begin{figure}[H]
	\centering
	\subfigure{\includegraphics[width=5cm]{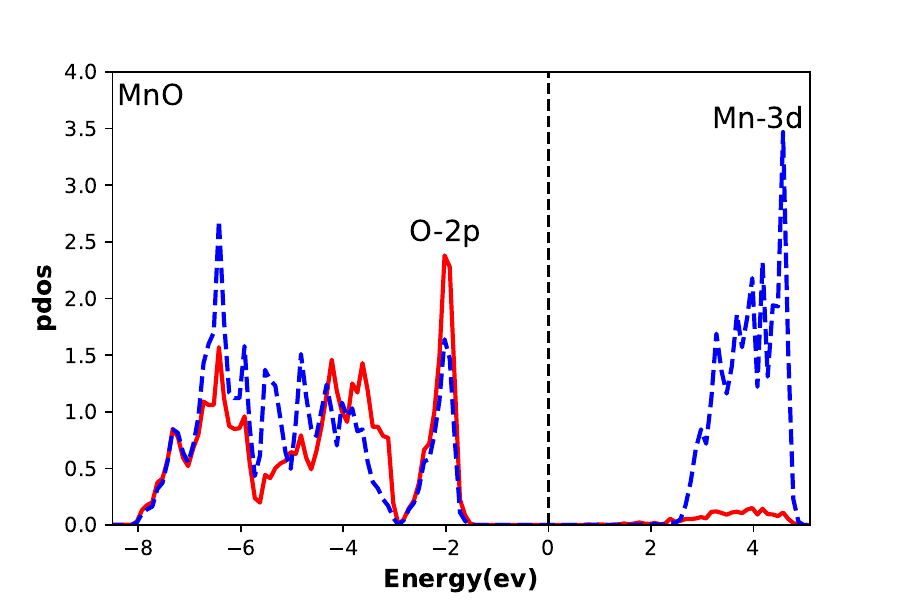}}
	\subfigure{\includegraphics[width=5cm]{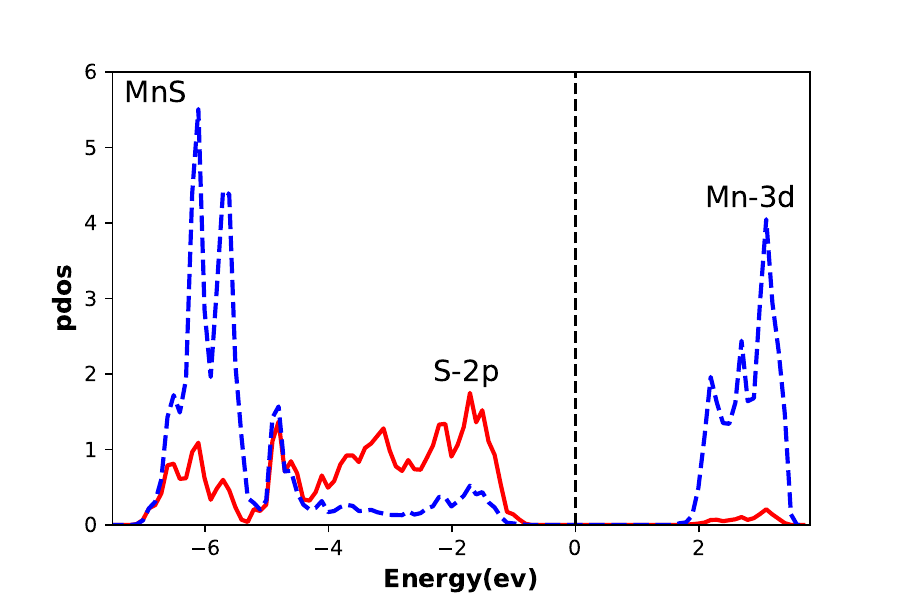}}
	\subfigure{\includegraphics[width=5cm]{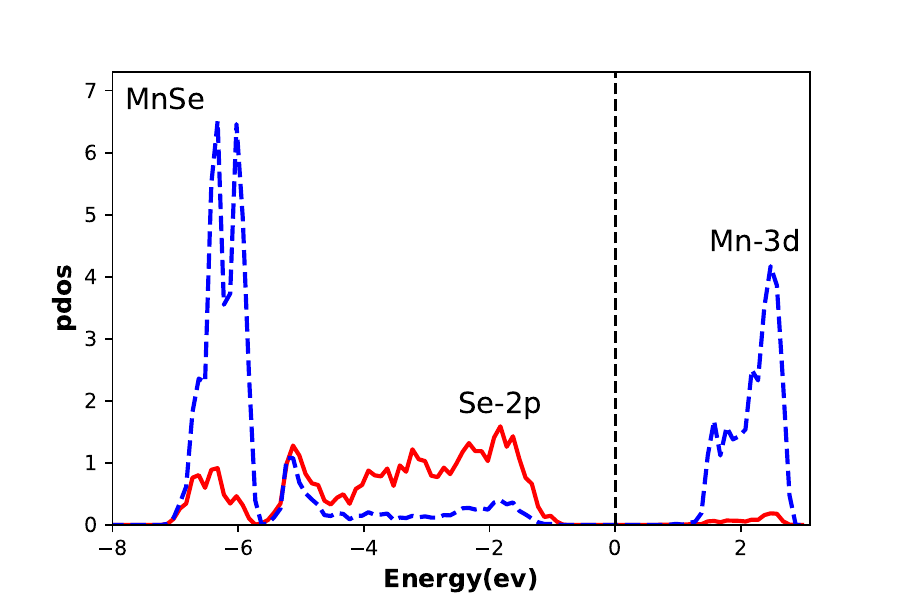}}
	\qquad
	\subfigure{\includegraphics[width=5cm]{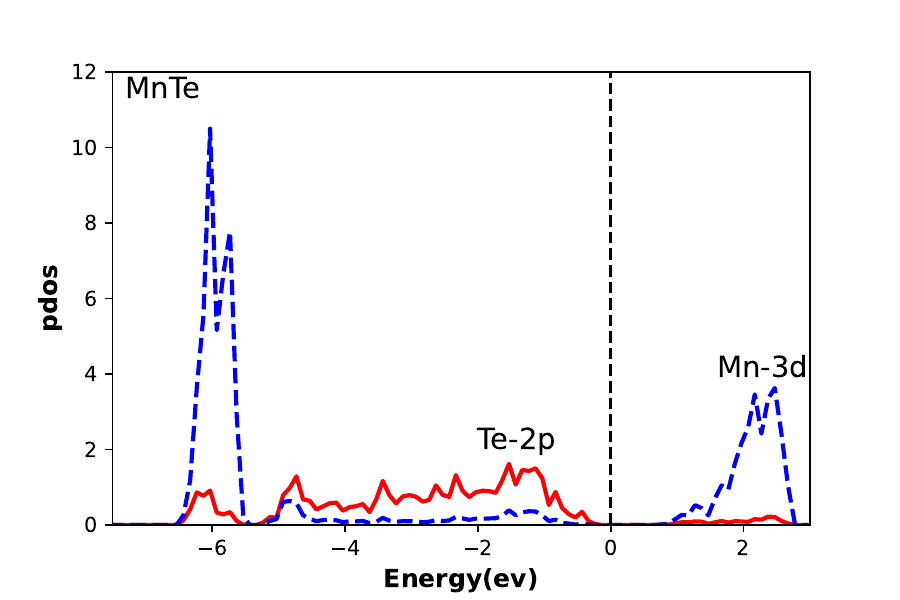}}
	\subfigure{\includegraphics[width=5cm]{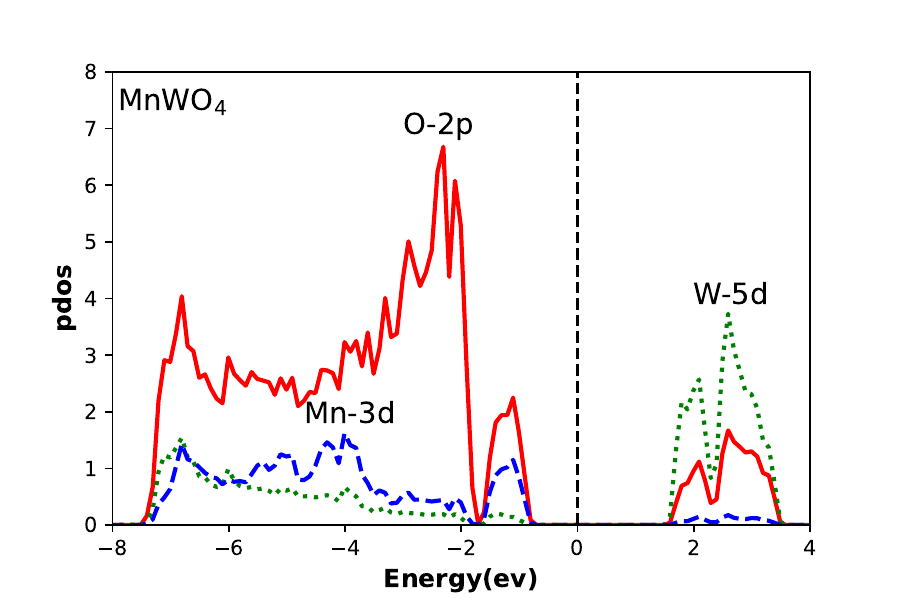}}
	\subfigure{\includegraphics[width=5cm]{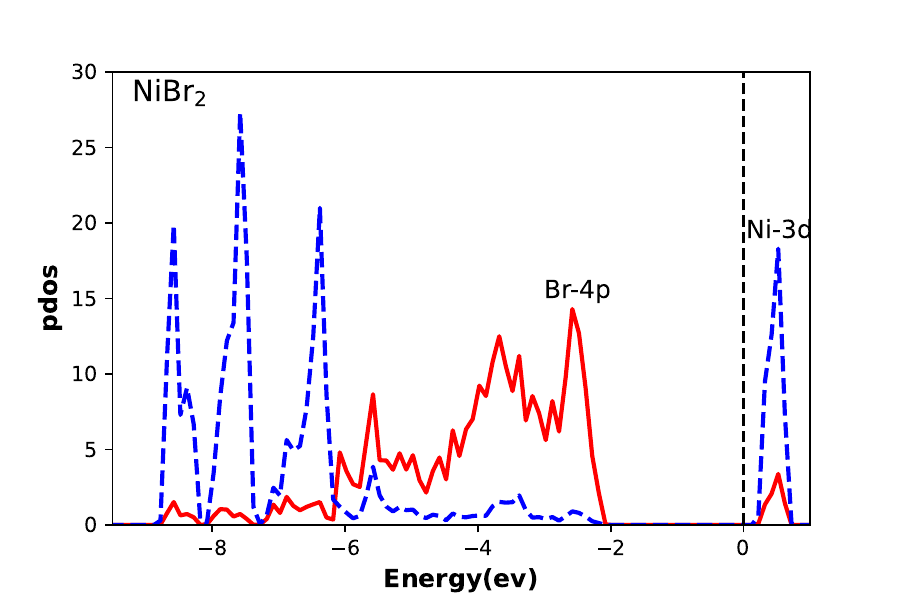}}
	\qquad
	\subfigure{\includegraphics[width=5cm]{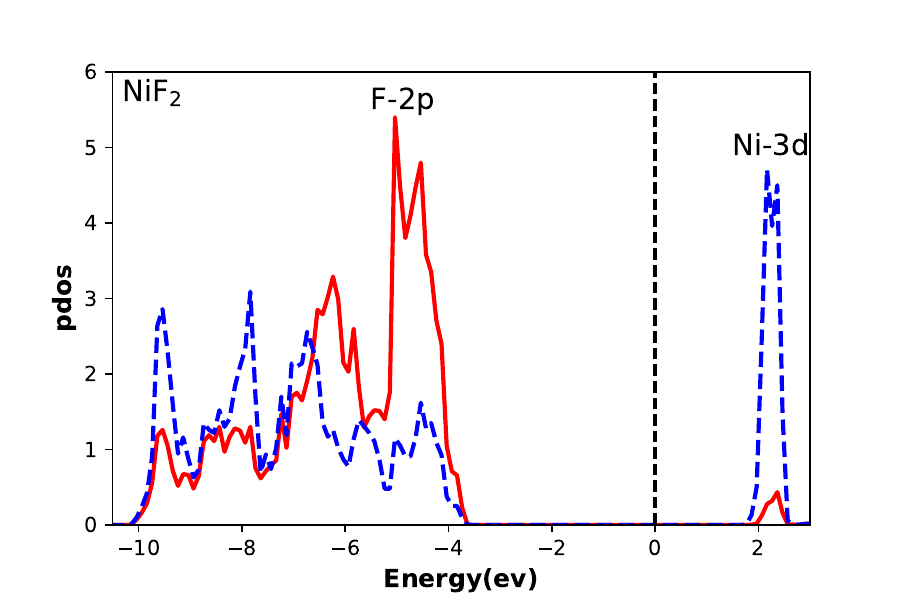}}
	\subfigure{\includegraphics[width=5cm]{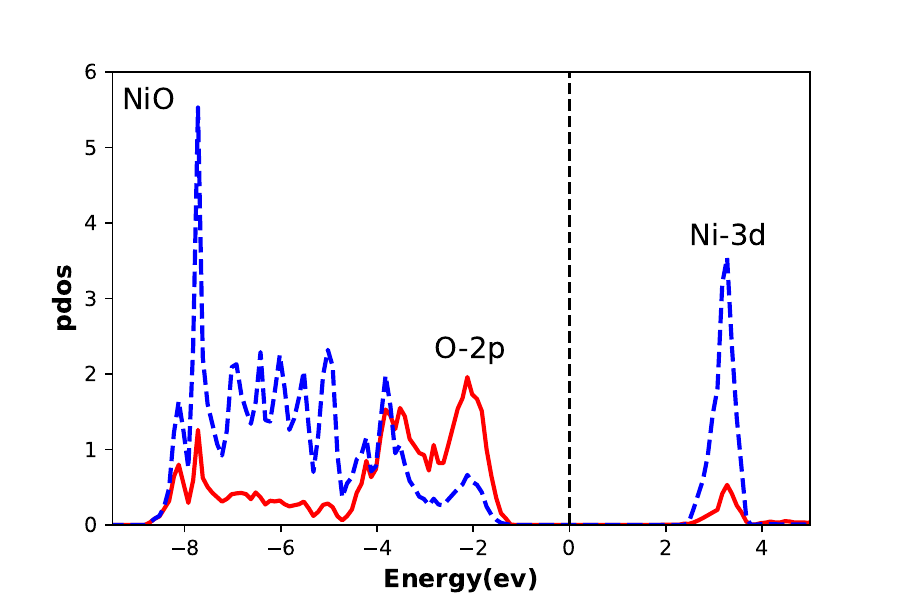}}
	\subfigure{\includegraphics[width=5cm]{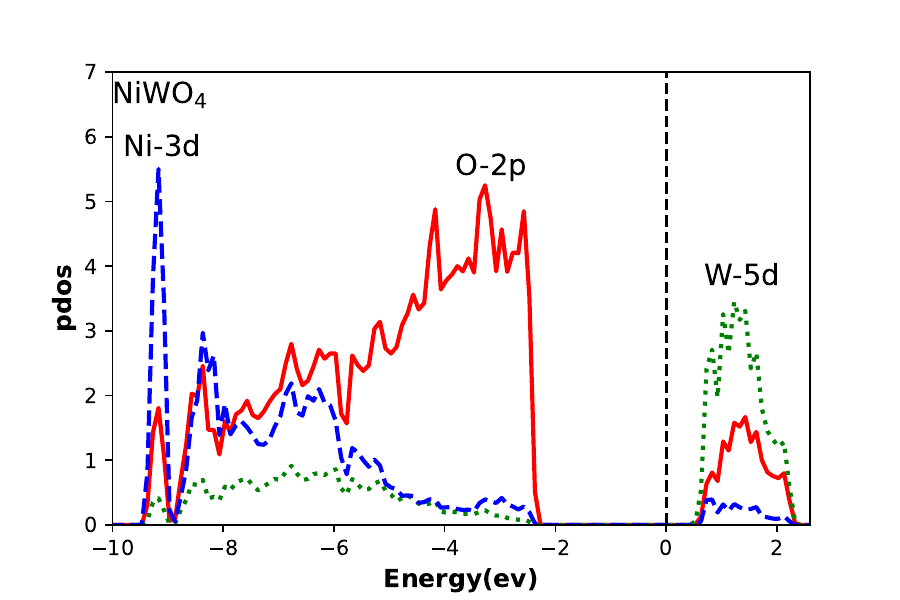}}
	\qquad
	\subfigure{\includegraphics[width=5cm]{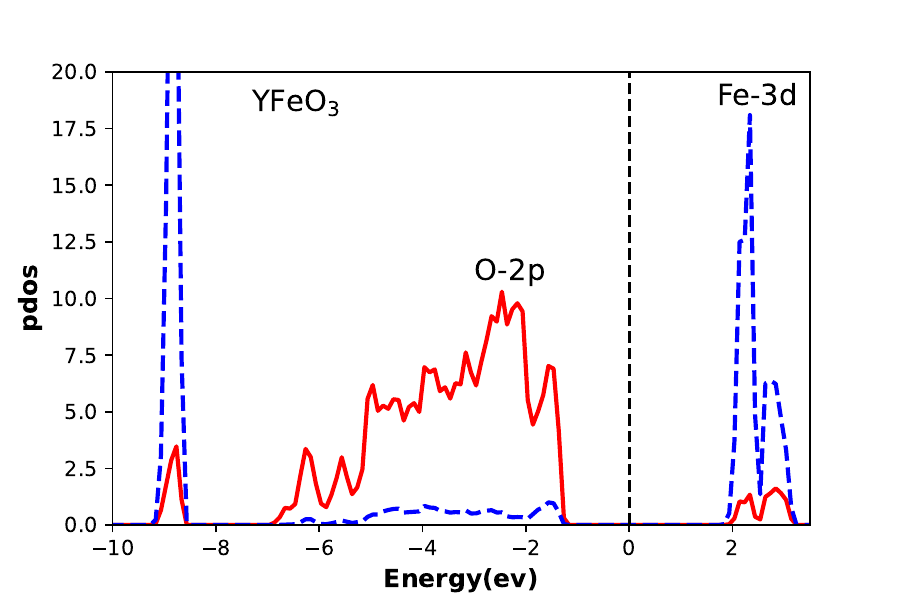}}
	\subfigure{\includegraphics[width=5cm]{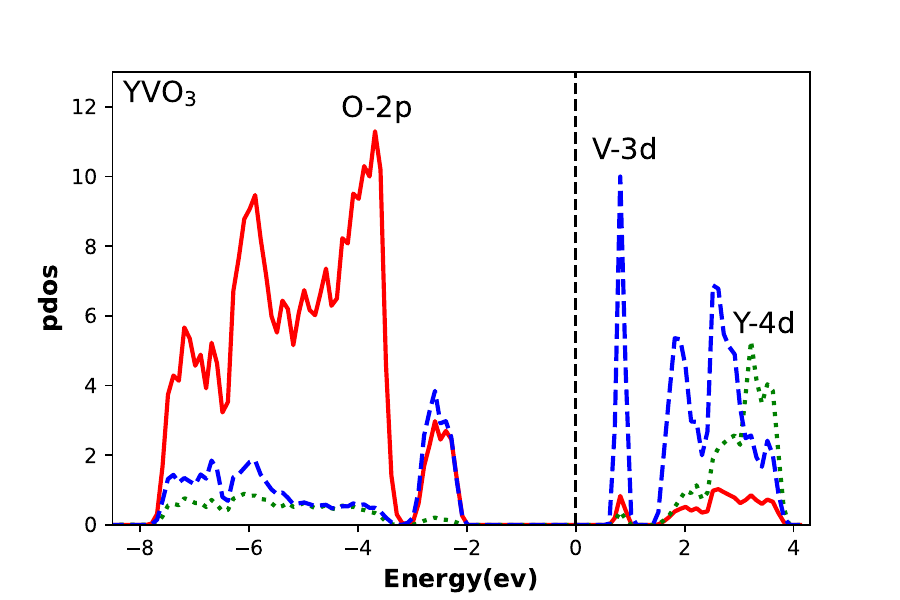}}  
	\caption{ (Color online) Projected density of states (PDOS) in  GGA+$U$ approximation. }
	\label{fig:S2}
\end{figure}

Here we present the project density of states (PDOS) for all compounds in Figures  \ref{fig:S1} and \ref{fig:S2} using GGA+$U$ calculations. 
For simplicity, orbitals with minimal contribution to the density of states (such as the 5$p$ orbital of the Te atom in the Cr$_2$TeO$_6$) are not included in the plots. 
\section{Heisenberg exchange parameters}
In this study, we map DFT results into Heisenberg Hamiltonian to derive exchange parameters ($J_{ij}$). 
For this aim, we calculate the DFT total energy of some magnetic configurations with a proper supercell. 
Then using the least squares method, we obtain the fittest values for exchange parameters. 
The following subsections are the results of exchange parameters from different approximations.
\subsection{Comparison between full potential and pseudopotential approaches}
Here we compare the full potential approach (using FPLO code) 
with the pseudopotential approach (using QE code) used in the paper to determine 
if the chosen pseudopotentials are appropriate to obtain exchange parameters.
We employed the GGA approximation (i.e., PBE) in both the QE package and FPLO code to carry out our calculations. 
However, due to the notable computational cost and time required for the FPLO code, 
we limited our analysis to only 12 configurations for all structures. 
Results are presented in Tables \ref{tab:fplo1}, \ref{tab:fplo2}, \ref{tab:fplo3} and \ref{tab:fplo4}. 
The results of the two methods demonstrate a notable level of agreement.
For some compounds (.i.e. LiCoPO$_4$, LaFeO$_3$ and La$_2$NiO$_4$), FPLO total energy does not converge for some configurations, so we don't present them in the table.
\begin{table}
	\centering
	\caption{Heisenberg exchange coefficient in QE and FPLO code. The table also provides information on both the distance between neighbors and the number of neighbors.}
	\begin{tabular}{lccccccccc}
		\hline
		Sample   & \makecell{$J_{GGA}$\\ (meV)} & \makecell{$J_{FPLO}$\\ (meV)}   & \makecell{$Distance$\\ ( \AA)}  & \hspace{5mm} \textbf{$Number\hspace{1mm} of\hspace{1mm} neighbours$}  \\
		\hline
		NiO    & $	\begin{aligned}
		&	J_1= -4.53\\
		&	J_2= -49.56\\
		&	J_3=  \hspace{3mm}5.51\\
		&	J_4= -2.22
		\end{aligned}$     & 
		$	\begin{aligned}
		&	J_1= \hspace{3mm} 2.89\\
		&	J_2=-51.24\\
		&	J_3=\hspace{3mm}1.67\\
		&	J_4=-4.81
		\end{aligned}$ & $	\begin{aligned}
		&	2.949342\\
		&	4.171000\\
		&	5.108411\\
		&	5.898685
		\end{aligned}$ & $	\begin{aligned}
		&	12\\
		&	6\\
		&	24\\
		&	12
		\end{aligned}$\\
		\hline
		MnO    & $	\begin{aligned}
		&	J_1=-7.52\\
		&	J_2= -16.99\\
		&	J_3= -0.21\\
		&	J_4= -2.43
		\end{aligned}$     & $	\begin{aligned}
		&	  J_1=-7.47\\
		&     J_2=-18.51\\
		&	  J_3= -0.03\\
		&	  J_4= -2.51
		\end{aligned}$ & $	\begin{aligned}
		&	3.143090\\
		&	4.445000\\
		&	5.443991\\
		&	6.286179
		\end{aligned}$ & $	\begin{aligned}
		&	12\\
		&	6\\
		&	24\\
		&	12
		\end{aligned}$\\
		\hline
		MnS    & $	\begin{aligned}
		&	J_1= -4.12\\
		&	J_2= -21.91\\
		&	J_3=\hspace{3mm}1.94\\
		&	J_4=-3.48
		\end{aligned}$     &  $	\begin{aligned}
		&	J_1=-0.33\\
		&	J_2= -22.02\\
		&  	J_3= -1.30\\
		&	J_4=-3.32
		\end{aligned}$ & $	\begin{aligned}
		&	3.691097\\
		&	5.220000\\
		&	6.393168\\
		&	7.382194
		\end{aligned}$ & $	\begin{aligned}
		&	12\\
		&	6\\
		&	24\\
		&	12
		\end{aligned}$\\
		\hline
		MnSe    & $	\begin{aligned}
		&	J_1= \hspace{3mm}1.07\\
		&	J_2=-19.62\\
		&	J_3= -0.69\\
		&	J_4= -4.22
		\end{aligned}$     & 
		$	\begin{aligned}
		&	J_1=  \hspace{3mm}0.83\\
		&	J_2= -21.15\\
		&	J_3= -1.61\\
		&	J_4= -3.98
		\end{aligned}$ & $	\begin{aligned}
		&	3.860801\\
		&	5.459998\\
		&	6.687104\\
		&	7.721603
		\end{aligned}$ & $	\begin{aligned}
		&	12\\
		&	6\\
		&	24\\
		&	12
		\end{aligned}$\\
		\hline
		Cr$_2$O$_3$    & $	\begin{aligned}
		&  J_1=-44.23\\
		&  J_2=-29.38\\
		&  J_3=\hspace{3mm}2.07\\
		&  J_4=\hspace{3mm}6.24\\
		&  J_5= -4.13\\
		&  J_6= -0.83\\
		&  J_7=-0.92\\
		&  J_8= -0.65\\
		&  J_9=-3.82
		\end{aligned}$     & $	\begin{aligned}
		&	J_1=-43.68\\
		&	J_2= -29.46\\
		&	J_3= \hspace{3mm}1.07\\
		&	J_4=\hspace{3mm}6.38\\
		&   J_5=-4.22\\
		&   J_6=  -0.92\\
		&   J_7=-0.94\\
		&   J_8=-0.96\\
		&   J_9=  -4.10
		\end{aligned}$ & $	\begin{aligned}
		&	2.651965\\
		&	2.890038\\
		&	3.426830\\
		&	3.652600\\
		&   4.147946\\
		&   4.961000\\
		&   5.362314\\
		&   5.625339\\
		&   5.691791
		\end{aligned}$ & $	\begin{aligned}
		&	1\\
		&	3 \\
		&	3  \\
		&	6\\
		&   1\\
		&  6\\
		&  6\\
		&  6\\
		&   3
		\end{aligned}$\\
		\hline
		Fe$_2$O$_3$    & $	\begin{aligned}
		&  J_1=\hspace{3mm}55.00\\
		&  J_2=-17.40\\
		&  J_3=-122.03\\
		&  J_4= \hspace{3mm}21.37\\
		&  J_5=\hspace{3mm}12.34\\
		&  J_6=\hspace{3mm}4.24\\
		&  J_7=\hspace{3mm}3.43\\
		&  J_8=-21.64\\
		&  J_9= \hspace{3mm}19.61
		\end{aligned}$     & 	$	\begin{aligned}
		&	J_1=\hspace{3mm}46.60\\
		&	J_2= -8.82\\
		&	J_3=-119.39\\
		&	J_4=\hspace{3mm}16.83\\
		&   J_5= \hspace{3mm}13.89\\
		&   J_6= \hspace{3mm}3.16\\
		&   J_7=\hspace{3mm}5.75\\
		&   J_8= -17.43\\
		&   J_9=\hspace{3mm}18.38
		\end{aligned}$ & $	\begin{aligned}
		&	2.827057\\
		&	2.955439\\
		&	3.398667\\
		&	3.703052\\
		&   4.054777\\
		&   5.035000\\
		&   5.431311\\
		&   5.774381\\
		&   5.838308
		\end{aligned}$ & $	\begin{aligned}
		&	1\\
		&	3 \\
		&	3  \\
		&	6\\
		&   1\\
		&  6\\
		&  6\\
		&  6\\
		&   3
		\end{aligned}$\\
		\hline
		BiFeO$_3$    & $	\begin{aligned}
		&	J_1= -63.23\\
		&	J_2= -1.26\\
		&	J_3=-1.84\\
		&	J_4=\hspace{3mm}0.25
		\end{aligned}$     & $	\begin{aligned}
		&	J_1= -63.41\\
		&	J_2=\hspace{3mm}2.31\\
		&	J_3= -1.42\\
		&	J_4=\hspace{3mm}2.31
		\end{aligned}$ & $	\begin{aligned}
		&	3.965448\\
		&	5.579999\\
		&	5.635844\\
		&	6.845522
		\end{aligned}$ & $	\begin{aligned}
		&	6\\
		&	6\\
		&	6\\
		&	6
		\end{aligned}$\\
		\hline
		NiBr$2$    & $	\begin{aligned}
		&  J_1=\hspace{3mm}7.86\\
		&  J_2=\hspace{3mm}0.30\\
		&  J_3=\hspace{3mm}0.00\\
		&  J_4= -3.22\\
		&  J_5= -1.66\\
		&  J_6= -0.24
		\end{aligned}$     & $	\begin{aligned}
		&	 J_1=\hspace{3mm}6.65\\
		&	 J_2=\hspace{3mm}0.22\\
		&	 J_3=\hspace{3mm}0.01\\
		&	 J_4=-2.86\\
		&    J_5=-1.48\\
		&    J_6=-0.22
		\end{aligned}$ & $	\begin{aligned}
		&	3.715538\\
		&	6.435501\\
		&	6.467034\\
		&	7.431077\\
		&   7.458401\\
		&   8.332645
		\end{aligned}$ & $	\begin{aligned}
		&	6\\
		&	6\\
		&	6\\
		&	6\\
		&   6\\
		&  12
		\end{aligned}$\\
		\hline
		YVO$3$    & $	\begin{aligned}
		&  J_1=\hspace{3mm}14.59\\
		&  J_2=\hspace{3mm}17.81\\
		&  J_3=\hspace{3mm}6.40\\
		&  J_4=-1.48\\
		&  J_5=-7.24\\
		&  J_6=-5.39\\
		&  J_7=\hspace{3mm}1.33
		\end{aligned}$     & $	\begin{aligned}
		&	J_1=\hspace{3mm}13.04\\
		&	J_2=\hspace{3mm}17.53\\
		&	J_3=\hspace{3mm}7.72\\
		&	J_4=-1.17\\
		&   J_5=-7.82\\
		&   J_6=-6.242\\
		&   J_7=\hspace{3mm}1.78
		\end{aligned}$  & $	\begin{aligned}
		&	3.775479\\
		&	3.845106\\
		&	5.282100\\
		&	5.388792\\
		&   5.589167\\
		&   6.492674\\
		&   6.744852
		\end{aligned}$ & $	\begin{aligned}
		&	2\\
		&	4\\
		&	2\\
		&	8\\
		&   2\\
		&   4\\
		&   4
		\end{aligned}$\\
		\hline
	\end{tabular}
	\label{tab:fplo1}
\end{table}
\begin{table}
	\centering
	\caption{Heisenberg exchange coefficient  in QE and FPLO code.  The table also provides information on both the distance between neighbors and the number of neighbors.}
	\begin{tabular}{lccccccccc}
		\hline
		Sample   & \makecell{$J_{GGA}$\\ (meV)} & \makecell{$J_{FPLO}$\\ (meV)}   & \makecell{$Distance$\\ ( \AA)}  & \hspace{5mm} \textbf{$Number\hspace{1mm} of\hspace{1mm} neighbours$}  \\
		\hline
		LiMnPO$4$    & $	\begin{aligned}
		&   J_1= -6.67\\
		&	J_2= -0.30\\
		&	J_3= -2.64\\
		&	J_4= -0.25\\
		&   J_5= -0.40\\
		&   J_6= -1.13
		\end{aligned}$     & $	\begin{aligned}
		&	 J_1= -7.17\\
		&	 J_2= -0.32\\
		&	 J_3=  -2.84\\
		&	 J_4= -0.24\\
		&    J_5=-0.44\\
		&    J_6= -1.27
		\end{aligned}$  & $	\begin{aligned}
		&	3.929788\\
		&	4.749769\\
		&	5.485724\\
		&	5.639180\\
		&   5.858933\\
		&   6.111644
		\end{aligned}$ & $	\begin{aligned}
		&	4\\
		&	2\\
		&	2\\
		&	2\\
		&   2\\
		&   2
		\end{aligned}$\\
		\hline
		LiNiPO$4$    & $	\begin{aligned}
		&   J_1=-4.60\\
		&	J_2=\hspace{3mm}0.41\\
		&	J_3=  -4.72\\
		&	J_4= -0.19\\
		&   J_5=\hspace{3mm}0.74\\
		&   J_6= -3.99
		\end{aligned}$     &  $	\begin{aligned}
		&	 J_1=  -5.58\\
		&	 J_2=\hspace{3mm}0.31\\
		&	 J_3= -4.55\\
		&	 J_4=-0.22\\
		&    J_5=\hspace{3mm}0.56\\
		&    J_6=-3.89
		\end{aligned}$  & $	\begin{aligned}
		&	3.781396\\
		&	4.676800\\
		&	5.372511\\
		&	5.467003\\
		&   5.605283\\
		&   5.853900
		\end{aligned}$ & $	\begin{aligned}
		&	4\\
		&	2\\
		&	2\\
		&	2\\
		&   2\\
		&   2
		\end{aligned}$\\
		\hline
		LiCoPO$4$    & $	\begin{aligned}
		&   J_1=\hspace{3mm}2.69\\
		&	J_2=\hspace{3mm}1.15\\
		&	J_3= -3.31\\
		&	J_4=\hspace{3mm}0.37\\
		&   J_5=\hspace{3mm}0.08\\
		&   J_6=-3.30
		\end{aligned}$     & $	\begin{aligned}
		&	 J_1= -  \\
		&	 J_2= - \\
		&	 J_3= - \\
		&	 J_4= - \\
		&        J_5= - \\
		&        J_6= - 
		\end{aligned}$  & $	\begin{aligned}
		&	3.824533\\
		&	4.708078\\
		&	5.408060\\
		&	5.536015\\
		&   5.704034\\
		&   5.916841
		\end{aligned}$ & $	\begin{aligned}
		&	4\\
		&	2\\
		&	2\\
		&	2\\
		&   2\\
		&   2
		\end{aligned}$\\
		\hline
		YFeO$3$    & $	\begin{aligned}
		&   J_1=-34.39\\
		&	J_2=-33.48\\
		&	J_3= -1.60\\
		&	J_4=\hspace{3mm}8.27\\
		&   J_5=\hspace{3mm}10.14\\
		&   J_6=-2.14\\
		&   J_7=-9.47
		\end{aligned}$     & $	\begin{aligned}
		&	 J_1=-29.60\\
		&	 J_2= -32.60\\
		&	 J_3=\hspace{3mm}2.67\\
		&	 J_4=\hspace{3mm}13.44\\
		&    J_5=\hspace{3mm}21.03\\
		&    J_6=-6.89\\
		&    J_7= -13.88
		\end{aligned}$  & $	\begin{aligned}
		&	3.797754\\
		&	3.842101\\
		&	5.274583\\
		&	5.402285\\
		&   5.588000\\
		&   6.499551\\
		&   6.756381
		\end{aligned}$ & $	\begin{aligned}
		&	2\\
		&	4\\
		&	2\\
		&	8\\
		&   2\\
		&   4\\
		&   4
		\end{aligned}$\\
		\hline
		LaFeO$3$    & $	\begin{aligned}
		&   J_1=  -33.35\\
		&	J_2=  -38.80\\
		&	J_3=-1.15\\
		&	J_4= -1.58\\
		&   J_5=\hspace{3mm}0.87\\
		&   J_6=\hspace{3mm}0.77		
		\end{aligned}$     & 	$	\begin{aligned}
		&	 J_1= -\\
		&	 J_2= - \\
		&	 J_3= -  \\
		&	 J_4= -   \\
		&        J_5= -  \\
		&        J_6= -  
		\end{aligned}$  & $	\begin{aligned}
		&	3.922913\\
		&	3.931164\\
		&	5.553674\\
		&	5.565004\\
		&   6.799718\\
		&   6.808709
		\end{aligned}$ & $	\begin{aligned}
		&	2\\
		&	4\\
		&	10\\
		&	2\\
		&   4\\
		&   4
		\end{aligned}$\\
		\hline
		LiMnO$2$    & $	\begin{aligned}
		&   J_1=-53.28\\
		&	J_2= -8.56\\
		&	J_3= -39.37\\
		&	J_4= -0.41\\
		&   J_5=  -0.71\\
		&   J_6=  -1.78
		\end{aligned}$     & $	\begin{aligned}
		&	 J_1= -54.58\\
		&	 J_2=-8.92\\
		&	 J_3=-40.37\\
		&	 J_4= -0.41\\
		&    J_5= -1.07\\
		&    J_6= -1.88
		\end{aligned}$  & $	\begin{aligned}
		&	2.804300\\
		&	3.135868\\
		&	4.547243\\
		&	4.894132\\
		&   5.055874\\
		&   5.342427
		\end{aligned}$ & $	\begin{aligned}
		&	2\\
		&	4\\
		&	2\\
		&	4\\
		&   4\\
		&   4
		\end{aligned}$\\
		\hline
		CrCl$2$    & $	\begin{aligned}
		&   J_1=-14.45\\
		&	J_2=\hspace{3mm}3.26\\
		&	J_3=\hspace{3mm}0.27\\
		&	J_4= -0.63\\
		&   J_5=-0.53\\
		&   J_6=\hspace{3mm}0.37
		\end{aligned}$     & $	\begin{aligned}
		&	 J_1=-13.80\\
		&	 J_2=\hspace{3mm}3.10\\
		&	 J_3=\hspace{3mm}0.29\\
		&	 J_4= -0.69\\
		&    J_5=  -0.58\\
		&    J_6=\hspace{3mm}0.35
		\end{aligned}$  & $	\begin{aligned}
		&	3.481577\\
		&	4.783220\\
		&	5.959699\\
		&	6.624000\\
		&   6.864543\\
		&   6.902129
		\end{aligned}$ & $	\begin{aligned}
		&	2\\
		&	8\\
		&	2\\
		&	2\\
		&   8\\
		&   4
		\end{aligned}$\\
		\hline
		KNiPO$_4$    & $	\begin{aligned}
		&	J_1=  -7.44\\
		&	J_2=\hspace{3mm}1.47\\
		&	J_3=\hspace{3mm}0.57\\
		&	J_4=-1.47
		\end{aligned}$     & $	\begin{aligned}
		&	 J_1= -8.28\\
		&	 J_2=\hspace{3mm}1.09\\
		&	 J_3=\hspace{3mm}0.30\\
		&	 J_4=  -1.53
		\end{aligned}$ & $	\begin{aligned}
		&	3.709826\\
		&	4.924296\\
		&	5.503407\\
		&	5.774551
		\end{aligned}$ & $	\begin{aligned}
		&	2\\
		&	2\\
		&	4\\
		&	2
		\end{aligned}$\\
		\hline
		MnF$_2$    & $	\begin{aligned}
		&	J_1= -2.98\\
		&	J_2= -7.51\\
		&	J_3= -0.09\\
		&	J_4= -0.20\\
		&   J_5=\hspace{3mm}0.05
		\end{aligned}$     & 	$	\begin{aligned}
		&	J_1= -3.33\\
		&	J_2= -7.96\\
		&	J_3= -0.09\\
		&	J_4= -0.20\\
		&   J_5=\hspace{3mm}0.05
		\end{aligned}$  & $	\begin{aligned}
		&	3.308455\\
		&	3.812046\\
		&	4.857000\\
		&	5.876761\\
		&   6.035184
		\end{aligned}$ & $	\begin{aligned}
		&	2\\
		&	8\\
		&	4\\
		&	8\\
		&   8
		\end{aligned}$\\
		\hline
	\end{tabular}
	\label{tab:fplo2}
\end{table}
\begin{table}
	\centering
	\caption{Heisenberg exchange coefficient in QE and FPLO code.  The table also provides information on both the distance between neighbors and the number of neighbors.}
	\begin{tabular}{lccccccccc}
		\hline
		Sample   & \makecell{$J_{GGA}$\\ (meV)} & \makecell{$J_{FPLO}$\\ (meV)}   & \makecell{$Distance$\\ ( \AA)}  & \hspace{5mm} \textbf{$Number\hspace{1mm} of\hspace{1mm} neighbours$}  \\
		\hline
		NiF$_2$    & $	\begin{aligned}
		&	J_1=-3.76\\
		&	J_2= -8.50\\
		&	J_3=-0.04\\
		&	J_4= -0.47\\
		&   J_5=\hspace{3mm}0.41
		\end{aligned}$     &  $	\begin{aligned}
		&	J_1= -3.80\\
		&	J_2= -8.43\\
		&	J_3= -0.03\\
		&	J_4=-0.43\\
		&   J_5=\hspace{3mm}0.35
		\end{aligned}$ & $	\begin{aligned}
		&	3.083600\\
		&	3.631391\\
		&	4.649700\\
		&	5.579274\\
		&   5.674873
		\end{aligned}$ & $	\begin{aligned}
		&	2\\
		&	8\\
		&	4\\
		&	8\\
		&   8
		\end{aligned}$\\
		\hline
		Fe$_2$TeO$_6$    & $	\begin{aligned}
		&	J_1=-12.60\\
		&	J_2= -32.95\\
		&	J_3=-1.47\\
		&	J_4=-5.02\\
		&   J_5=\hspace{3mm}1.69
		\end{aligned}$     & 	$	\begin{aligned}
		&	J_1= -4.43\\
		&	J_2= -34.06\\
		&	J_3=\hspace{3mm}3.58\\
		&	J_4= -11.54\\
		&   J_5=\hspace{3mm}5.54
		\end{aligned}$  & $	\begin{aligned}
		&	3.031173\\
		&	3.591210\\
		&	4.604300\\
		&	5.512494\\
		&   5.592209
		\end{aligned}$ & $	\begin{aligned}
		&	1\\
		&   4\\
		&	4\\
		&	4\\
		&   8
		\end{aligned}$\\
		\hline
		La$_2$NiO$_4$    & $	\begin{aligned}
		&	J_1= -38.12 \\
		&	J_2=\hspace{3mm}7.32 \\
		&	J_3=\hspace{3mm}0.04 
		\end{aligned}$     & 	$	\begin{aligned}
		&	J_1= - \\
		&	J_2= - \\
		&	J_3= - 
		\end{aligned}$  & $	\begin{aligned}
		&	3.850000\\
		&	5.444722\\
		&   6.811405
		\end{aligned}$ & $	\begin{aligned}
		&	4\\
		&	4\\
		&	8
		\end{aligned}$\\
		\hline
		Cr$_2$TeO$_6$   & $	\begin{aligned}
		&	J_1=-26.48\\
		&	J_2=-11.99\\
		&	J_3= -1.91\\
		&	J_4=\hspace{3mm}0.85\\
		&   J_5=-1.18
		\end{aligned}$     & 	$	\begin{aligned}
		&	J_1=-25.14\\
		&	J_2=-12.12\\
		&	J_3=-1.90\\
		&	J_4=\hspace{3mm}0.64\\
		&   J_5= -1.13
		\end{aligned}$  & $	\begin{aligned}
		&	2.983634\\
		&	3.557204\\
		&	4.546000\\
		&	5.437664\\
		&   5.535893
		\end{aligned}$ & $	\begin{aligned}
		&	1\\
		&	4\\
		&	4\\
		&	4\\
		&   8
		\end{aligned}$\\
		\hline
		KMnSb    & $	\begin{aligned}
		&	J_1= -82.74\\
		&	J_2=-11.10\\
		&	J_3=\hspace{3mm}6.44\\
		&	J_4=\hspace{3mm}3.52\\
		&   J_5=\hspace{3mm}7.40\\
		&   J_6=-4.50\\
		&   J_7=-11.10
		\end{aligned}$     & $	\begin{aligned}
		&	J_1= -77.78\\
		&	J_2=-12.71\\
		&	J_3=\hspace{3mm}7.93\\
		&	J_4=\hspace{3mm}3.02\\
		&   J_5=\hspace{3mm}9.89\\
		&   J_6=-6.75\\
		&   J_7= -12.71
		\end{aligned}$  & $	\begin{aligned}
		&	3.273197\\
		&	4.629000\\
		&	6.546394\\
		&	7.319092\\
		&   8.121000\\
		&   8.755824\\
		&   9.258000
		\end{aligned}$ & $	\begin{aligned}
		&  4\\
		&  4\\
		&  4\\
		&  8\\
		&  2\\
		&  8\\
		&  4
		\end{aligned}$\\
		\hline
		Cr$_2$WO$_6$    & $	\begin{aligned}
		&	J_1=  -7.68\\
		&	J_2=\hspace{3mm}2.34\\
		&	J_3=-2.98\\
		&	J_4=\hspace{3mm}5.26\\
		&   J_5=  -6.86
		\end{aligned}$     & $	\begin{aligned}
		&	J_1=  -7.02\\
		&	J_2=\hspace{3mm}1.44\\
		&	J_3= -3.49\\
		&	J_4=\hspace{3mm}6.09\\
		&   J_5=   -7.88
		\end{aligned}$  & $	\begin{aligned}
		&	2.934689\\
		&	3.568379\\
		&	4.580000\\
		&	5.439559\\
		&   5.490015
		\end{aligned}$ & $	\begin{aligned}
		&	1\\
		&	4\\
		&	4\\
		&	4\\
		&   8\
		\end{aligned}$\\
		\hline
		NiWo$_4$    & $	\begin{aligned}
		&	J_1=\hspace{3mm}1.53\\
		&	J_2=\hspace{3mm}6.80\\
		&	J_3= -1.29\\
		&	J_4=-5.99\\
		&   J_5=-16.59
		\end{aligned}$     & $	\begin{aligned}
		&	J_1=\hspace{3mm}1.60\\
		&	J_2=\hspace{3mm}6.61\\
		&	J_3= -1.04\\
		&	J_4= -6.57\\
		&   J_5= -17.03
		\end{aligned}$  & $	\begin{aligned}
		&	3.060435\\
		&	4.549336\\
		&	4.599200\\
		&	4.906800\\
		&   5.523322
		\end{aligned}$ & $	\begin{aligned}
		&  2\\
		&  2\\
		&  2\\
		&  2\\
		&  2
		\end{aligned}$\\
		\hline
		MnWO$_4$    & $	\begin{aligned}
		&	J_1= -14.40\\
		&	J_2=\hspace{3mm}1.49\\
		&	J_3= -2.57\\
		&	J_4=-7.90\\
		&   J_5=-19.62
		\end{aligned}$     & $	\begin{aligned}
		&	J_1= -13.63\\
		&	J_2=\hspace{3mm}1.71\\
		&	J_3= -2.77\\
		&	J_4=  -9.01\\
		&   J_5=  -22.43
		\end{aligned}$  & $	\begin{aligned}
		&	3.052166\\
		&	4.704916\\
		&	4.820000\\
		&	4.988685\\
		&   5.664495
		\end{aligned}$ & $	\begin{aligned}
		&  2\\
		&  2\\
		&  2\\
		&  2\\
		&  2
		\end{aligned}$\\
		\hline
		CoWO$_4$    & $	\begin{aligned}
		&	J_1=\hspace{3mm}5.96\\
		&	J_2=-0.32\\
		&	J_3=\hspace{3mm}2.01\\
		&	J_4=-1.71\\
		&   J_5=-0.66\\
		&   J_6=\hspace{3mm}1.01\\
		&   J_7=-1.25\\
		&   J_8=-0.64\\
		&   J_9= -0.74
		\end{aligned}$     & $	\begin{aligned}
		&	J_1= -\\
		&	J_2= -\\
		&	J_3= -\\
		&	J_4= -\\
		&       J_5= -\\
		&       J_6= - \\
		&       J_7= -\\
		&       J_8= -\\
		&       J_9= - 
		\end{aligned}$  & $	\begin{aligned}
		&	3.149966\\
		&	4.485357\\
		&	4.670000\\
		&	4.9517122\\
		&   5.633044\\
		&   5.687543\\
		&   6.475131\\
		&   6.806493\\
		&   7.359147
		\end{aligned}$ & $	\begin{aligned}
		&  2\\
		&  2\\
		&  2\\
		&  2\\
		&  4\\
		&  4\\
		&  4\\
		&  4\\
		&  4
		\end{aligned}$\\
		\hline
	\end{tabular}
	\label{tab:fplo3}
\end{table}
\begin{table}
	\centering
	\caption{Heisenberg exchange coefficient  in QE and FPLO code.  The table also provides information on both the distance between neighbors and the number of neighbors.}
	\begin{tabular}{ccccccccc}
		\hline
		Sample   & \makecell{$J_{GGA}$\\ (meV)} & \makecell{$J_{FPLO}$\\ (meV)}   & \makecell{$Distance$\\ ( \AA)}  & \hspace{5mm} \textbf{$Number\hspace{1mm} of\hspace{1mm} neighbours$}  \\
		\hline
		Li$_2$MnO$_3$   & $	\begin{aligned}
		&	J_1= -4.48\\
		&	J_2= -4.99\\
		&	J_3=\hspace{3mm}0.58\\
		&	J_4= \hspace{3mm}0.71\\
		&   J_5=  -0.06\\
		&   J_6= -0.13\\
		&   J_7= -0.92\\
		&   J_8= -0.94\\
		&   J_9=   -0.51
		\end{aligned}$     & $	\begin{aligned}
		&	J_1=  -5.67\\
		&	J_2=  -5.79\\
		&	J_3=\hspace{3mm}0.44\\
		&	J_4=\hspace{3mm}0.53\\
		&   J_5= -0.06\\
		&   J_6=-0.17\\
		&   J_7= -0.87\\
		&   J_8= -0.94\\
		&   J_9=  -0.47
		\end{aligned}$  & $	\begin{aligned}
		&	2.845271\\
		&	2.851053\\
		&	4.928717\\
		&	4.937000\\
		&   5.012320\\
		&   5.030000\\
		&   5.680947\\
		&   5.701094\\
		&   5.755748
		\end{aligned}$ & $	\begin{aligned}
		&  2\\
		&  2\\
		&  4\\
		&  2\\
		&  2\\
		&  2\\
		&  1\\
		&  2\\
		&  2
		\end{aligned}$\\
		\hline
		MnTe    & $	\begin{aligned}
		&	J_1= -46.07\\
		&	J_2= -1.45\\
		&	J_3=  -8.84\\
		&	J_4=\hspace{3mm}5.24\\
		&   J_5= -1.19\\
		&   J_6= -1.96\\
		&   J_7=\hspace{3mm}1.09
		\end{aligned}$     & $	\begin{aligned}
		&	J_1= -47.25\\
		&	J_2= -2.28\\
		&	J_3= -9.57\\
		&	J_4=\hspace{3mm}5.65\\
		&   J_5= -1.23\\
		&   J_6= -2.13\\
		&   J_7=\hspace{3mm}1.08
		\end{aligned}$  & $	\begin{aligned}
		&	3.351462\\
		&	4.150000\\
		&	5.334304\\
		&	6.702924\\
		&   7.188010\\
		&   7.883634\\
		&   7.930939
		\end{aligned}$ & $	\begin{aligned}
		&  2\\
		&  6\\
		&  12\\
		&  2\\
		& 6\\
		& 12\\
		& 12
		\end{aligned}$\\
		\hline
	\end{tabular}
	\label{tab:fplo4}
\end{table}
\noindent
\subsection{Heisenberg exchange paramters from GGA and GGA+$U$ approximation}
This part presents the Heisenberg exchange parameter derived from GGA and GGA+$U$ calculations in Tables \ref{tab:Hes1}, \ref{tab:Hes2}, \ref{tab:Hes3} and \ref{tab:Hes4}. 
For each compound, we use three times more than the minimal magnetic configurations needed to obtain the exchange parameter from the least squares method. 
For example, for NiO, the minimal magnetic configuration is 5 to derive the four exchange parameters. Here instead, we use 15 magnetic configurations.

\begin{table}
	\centering
	\caption{ Hesineberg exchanges  and transition temperatures   in GGA and GGA+$U$ approximation.}
	\label{tab:Hes1}
	\begin{tabular}{lcccc}
		\hline
		Sample     & \makecell{$J_{GGA}$\\ (meV)}   & \makecell{$J_{GGA+U}$\\ (meV)}  & \makecell{$T_{{GGA}}$\\ (K)} & \makecell{$T_{{GGA+U}}$\\ (K)}  \\
		\hline
		NiO    & $	\begin{aligned}
		&	J_1=-4.53\\
		&	J_2=-49.55\\
		&	J_3=\hspace{3mm}5.51\\
		&	J_4=-2.22
		\end{aligned}$     & 
		$	\begin{aligned}
		&	J_1=\hspace{3mm}2.22\\
		&	J_2=-15.26\\
		&	J_3=-0.54\\
		&	J_4= -0.70
		\end{aligned}$ & 695 & 220 \\
		\hline
		MnO   & $	\begin{aligned}
		&	J_1=-7.52\\
		&	J_2=-16.99\\
		&	J_3=-0.21\\
		&	J_4= -2.43
		\end{aligned}$     & 	$	\begin{aligned}
		&	J_1= -6.46\\
		&	J_2=-4.32\\
		&	J_3=-0.18\\
		&	J_4= -0.30
		\end{aligned}$ & 151 & 26 \\
		\hline
		MnS     & $	\begin{aligned}
		&	J_1= -4.12\\
		&	J_2=-21.91\\
		&	J_3=\hspace{3mm}1.94\\
		&	J_4= -3.48
		\end{aligned}$     & 	$	\begin{aligned}
		&	J_1=-1.02\\
		&	J_2= -5.65\\
		&  	J_3=-0.69\\
		&	J_4= -0.82
		\end{aligned}$ & 185 & 54 \\
		\hline
		MnSe    & $	\begin{aligned}
		&	J_1=\hspace{3mm}1.07\\
		&	J_2=-19.62\\
		&	J_3=-0.69\\
		&	J_4=-4.22
		\end{aligned}$     & 	$	\begin{aligned}
		&	J_1=-0.11\\
		&	J_2=-4.71\\
		&	J_3=-0.79\\
		&	J_4=-1.00
		\end{aligned}$  & 131 & 25 \\
		\hline
		Cr$_2$O$_3$    & $	\begin{aligned}
		&  J_1=-45.49\\
		&  J_2= -29.24\\
		&  J_3=\hspace{3mm}1.43\\
		&  J_4=\hspace{3mm}3.60\\
		&  J_5=-2.57\\
		&  J_6=\hspace{3mm}0.71\\
		&  J_7= -0.92\\
		&  J_8=-0.94\\
		&  J_9= -1.29
		\end{aligned}$     & 	$	\begin{aligned}
		&	J_1=-10.96\\
		&	J_2= -7.90\\
		&	J_3=\hspace{3mm}4.02\\
		&	J_4=\hspace{3mm}4.14\\
		&   J_5=\hspace{3mm}0.12\\
		&   J_6=\hspace{3mm}0.06\\
		&   J_7=\hspace{3mm}0.01\\
		&   J_8=\hspace{3mm}0.04\\
		&   J_9= -0.22
		\end{aligned}$ & 428 & 118  \\
		\hline
		Fe$_2$O$_3$    & $	\begin{aligned}
		&  J_1=\hspace{3mm}29.94\\
		&  J_2=\hspace{3mm}8.61\\
		&  J_3=-81.50\\
		&  J_4= -26.14\\
		&  J_5=-0.87\\
		&  J_6=\hspace{3mm}9.08\\
		&  J_7=-3.90\\
		&  J_8= -3.12\\
		&  J_9=\hspace{3mm}15.01
		\end{aligned}$     & 	$	\begin{aligned}
		&	J_1=\hspace{3mm}3.61\\
		&	J_2= -4.35\\
		&	J_3= -66.03\\
		&	J_4= -15.95\\
		&   J_5= -6.48\\
		&   J_6= -0.17\\
		&   J_7= -0.96\\
		&   J_8= -1.03\\
		&   J_9=-1.56
		\end{aligned}$ & 1446 & 701 \\
		\hline
		BiFeO$_3$    & $	\begin{aligned}
		&	J_1=-63.23\\
		&	J_2= -1.26\\
		&	J_3=-1.84\\
		&	J_4=\hspace{3mm}0.25
		\end{aligned}$     & $	\begin{aligned}
		&	J_1=-31.08\\
		&	J_2=-2.00\\
		&	J_3=-0.90\\
		&	J_4= -0.68
		\end{aligned}$ & 972 & 466 \\
		\hline
		NiBr$2$   & $	\begin{aligned}
		&  J_1=\hspace{3mm}7.87\\
		&  J_2=\hspace{3mm}0.32\\
		&  J_3=\hspace{3mm}0.02\\
		&  J_4=-3.20\\
		&  J_5=-1.65\\
		&  J_6= -0.29
		\end{aligned}$     & $	\begin{aligned}
		&  J_1=\hspace{3mm}1.98\\
		&  J_2=\hspace{3mm}0.05\\
		&  J_3=\hspace{3mm}0.02\\
		&  J_4=-0.82\\
		&  J_5=-0.36\\
		&  J_6=-0.05
		\end{aligned}$  & 93  & 19 \\
		\hline
		YVO$3$   & $	\begin{aligned}
		&  J_1=\hspace{3mm}13.29\\
		&  J_2=\hspace{3mm}17.70\\
		&  J_3= -1.68\\
		&  J_4=-1.01\\
		&  J_5=-1.38\\
		&  J_6=-0.84\\
		&  J_7=-2.51
		\end{aligned}$     & $	\begin{aligned}
		&	J_1= -1.92\\
		&	J_2=-4.08\\
		&	J_3= -0.03\\
		&	J_4= -0.06\\
		&   J_5= -0.32\\
		&   J_6=\hspace{3mm}0.01\\
		&   J_7=\hspace{3mm}0.03
		\end{aligned}$  & 142 & $49.2$\\
		\hline
	\end{tabular}
\end{table}
\begin{table}
	\centering
	\caption{ Hesineberg exchanges and transition temperatures   in GGA and GGA+$U$ approximation.}
	\label{tab:Hes2}
	\begin{tabular}{lcccc}
		\hline
		Sample     & \makecell{$J_{GGA}$\\ (meV)}   & \makecell{$J_{GGA+U}$\\ (meV)}  & \makecell{$T_{{GGA}}$\\ (K)} & \makecell{$T_{{GGA+U}}$\\ (K)}  \\
		\hline
		LiMnPO$4$   & $	\begin{aligned}
		&   J_1= -6.68\\
		&	J_2=-0.30\\
		&	J_3= -2.64\\
		&	J_4=-0.25\\
		&   J_5= -0.40\\
		&   J_6=-1.14
		\end{aligned}$     & 	$	\begin{aligned}
		&	 J_1= -2.29\\
		&	 J_2= -0.09\\
		&	 J_3= -0.95\\
		&	 J_4=-0.11\\
		&    J_5= -0.11\\
		&    J_6= -0.31
		\end{aligned}$ & 69 & 25\\
		\hline
		LiNiPO$4$    & $	\begin{aligned}
		&   J1=  -4.58\\
		&	J2=\hspace{3mm}0.44\\
		&	J3= -4.72\\
		&	J4= -0.20\\
		&   J5=\hspace{3mm}0.71\\
		&   J6=-3.98  
		\end{aligned}$     &  $	\begin{aligned}
		&	 J1= -1.25\\
		&	 J2= -0.02\\
		&	 J3= -0.85\\
		&	 J4=\hspace{3mm}0.06\\
		&    J5=\hspace{3mm}0.13\\
		&    J6=-1.05
		\end{aligned}$ & 48 & 11 \\
		\hline
		LiCoPO$4$    & $	\begin{aligned}
		&   J1=\hspace{3mm}2.69\\
		&	J2=\hspace{3mm}1.15\\
		&	J3= -3.31\\
		&	J4=\hspace{3mm}0.37\\
		&   J5=\hspace{3mm}0.08\\
		&   J6= -3.30
		\end{aligned}$     & $	\begin{aligned}
		&	 J1= -1.83\\
		&	 J2=-0.15\\
		&	 J3= -0.58\\
		&	 J4=-0.05\\
		&    J5=-0.09\\
		&    J6=-0.73 
		\end{aligned}$ & 27 & 14 \\
		\hline
		YFeO$3$   & $	\begin{aligned}
		&   J1= -32.87\\
		&	J2=-31.59\\
		&	J3=\hspace{3mm}2.30\\
		&	J4=\hspace{3mm}2.45\\
		&   J5= -3.33\\
		&   J6=\hspace{3mm}0.61\\
		&   J7= -1.18
		\end{aligned}$     & $	\begin{aligned}
		&	 J1=-29.34\\
		&	 J2=-32.25\\
		&	 J3=-0.33\\
		&	 J4= -0.91\\
		&    J5= -1.33\\
		&    J6=\hspace{3mm}0.16\\
		&    J7=\hspace{3mm}0.31
		\end{aligned}$ & 603 & 469 \\
		\hline
		LaFeO$3$   & $	\begin{aligned}
		&   J_1=  -33.35\\
		&	J_2=  -38.80\\
		&	J_3=-1.15\\
		&	J_4= -1.58\\
		&   J_5=\hspace{3mm}0.87\\
		&   J_6=\hspace{3mm}0.77
		\end{aligned}$     & 	$	\begin{aligned}
		&	 J_1= -11.13\\
		&	 J_2= -13.27\\
		&	 J_3=\hspace{3mm}8.95\\
		&	 J_4=\hspace{3mm}4.60\\
		&    J_5=-5.12\\
		&    J_6= -2.91
		\end{aligned}$ & 670 & 527 \\
		\hline
		LiMnO$2$   & $	\begin{aligned}
		&   J_1=-53.28\\
		&	J_2= -8.64\\
		&	J_3=-39.13\\
		&	J_4=-0.43\\
		&   J_5=-0.62\\
		&   J_6=-1.86
		\end{aligned}$     & $	\begin{aligned}
		&	 J_1= -17.08\\
		&	 J_2= -1.54\\
		&	 J_3= -13.59\\
		&	 J_4=  -0.03\\
		&    J_5=\hspace{3mm}0.73\\
		&    J_6= -0.51
		\end{aligned}$ & 362 & 111  \\
		\hline
		CrCl$2$   & $	\begin{aligned}
		&   J_1=-13.78\\
		&	J_2=\hspace{3mm}3.02\\
		&	J_3= -0.17\\
		&	J_4=  -0.33\\
		&   J_5= -0.32\\
		&   J_6=\hspace{3mm}0.05
		\end{aligned}$     & $	\begin{aligned}
		&	 J_1= -1.90\\
		&	 J_2=\hspace{3mm}0.29\\
		&	 J_3= -0.03\\
		&	 J_4= -0.03\\
		&    J_5=\hspace{3mm}0.02\\
		&    J_6=\hspace{3mm}0.01
		\end{aligned}$ & 35 & 2  \\
		\hline
		KNiPO$_4$  & $	\begin{aligned}
		&	J_1= -7.44\\
		&	J_2=\hspace{3mm}1.47\\
		&	J_3=\hspace{3mm}0.57\\
		&	J_4=-1.47
		\end{aligned}$     & 
		$	\begin{aligned}
		&	 J_1= -1.55\\
		&	 J_2=\hspace{3mm}0.17\\
		&	 J_3=\hspace{3mm}0.03\\
		&	 J_4= -0.33 
		\end{aligned}$ &53 & 9  \\
		\hline
		MnF$_2$    & $	\begin{aligned}
		&	J_1=  -2.84\\
		&	J_2=-7.46\\
		&	J_3= -0.15\\
		&	J_4=-0.20\\
		&   J_5=-0.01
		\end{aligned}$     & $	\begin{aligned}
		&	J_1=-0.44\\
		&	J_2= -2.72\\
		&	J_3=-0.04\\
		&	J_4= -0.07\\
		&   J_5=\hspace{3mm}0.02
		\end{aligned}$ & 146 & 57 \\
		\hline
	\end{tabular}
\end{table}
\begin{table}[H]
	\centering
	\caption{ Hesineberg exchanges  and transition temperatures   in GGA and GGA+$U$ approximation.}
	\label{tab:Hes3}
	\begin{tabular}{lcccc}
		\hline
		Sample     & \makecell{$J_{GGA}$\\ (meV)}   & \makecell{$J_{GGA+U}$\\ (meV)}  & \makecell{$T_{{GGA}}$\\ (K)} & \makecell{$T_{{GGA+U}}$\\ (K)}  \\
		\hline
		NiF$_2$   & $	\begin{aligned}
		&	J_1=-3.38\\
		&	J_2= -8.41\\
		&	J_3= -0.18\\
		&	J_4=-0.52\\
		&   J_5=\hspace{3mm}0.33
		\end{aligned}$     & $	\begin{aligned}
		&	J_1=-0.47\\
		&	J_2=-2.08\\
		&	J_3=-0.02\\
		&	J_4=-0.11\\
		&   J_5= \hspace{3mm}0.09
		\end{aligned}$ & 134  & 37  \\
		\hline
		Fe$_2$TeO$_6$   & $	\begin{aligned}
		&	J_1=-12.22\\
		&	J_2=-32.93\\
		&	J_3=-1.44\\
		&	J_4=-4.75\\
		&   J_5=\hspace{3mm}1.50
		\end{aligned}$     & 	$	\begin{aligned}
		&	J_1=-7.43\\
		&	J_2=-19.76\\
		&	J_3=-3.10\\
		&	J_4=\hspace{3mm}4.76\\
		&   J_5= -4.81
		\end{aligned}$  & 478 & 351 \\
		\hline
		La$_2$NiO$_4$  & $	\begin{aligned}
		&	J_1= -38.12 \\
		&	J_2=\hspace{3mm}7.32 \\
		&	J_3=\hspace{3mm}0.04
		\end{aligned}$     & 	$	\begin{aligned}
		&	J_1= -26.99\\
		&	J_2= -1.27\\
		&	J_3= \hspace{3mm}0.01
		\end{aligned}$ & 395& 209 \\
		\hline
		Cr$_2$TeO$_6$   & $	\begin{aligned}
		&	J_1=-26.90\\
		&	J_2= -12.01\\
		&	J_3=-1.94\\
		&	J_4=\hspace{3mm}0.75\\
		&   J_5=-1.08
		\end{aligned}$     & 	$	\begin{aligned}
		&	J_1=-4.47\\
		&	J_2= -1.98\\
		&	J_3=-0.63\\
		&	J_4=\hspace{3mm}0.69\\
		&   J_5=-0.24
		\end{aligned}$ & 106 &  17 \\
		\hline
		KMnSb     & $	\begin{aligned}
		&	J_1= -88.07\\
		&	J_2= -9.68\\
		&	J_3= \hspace{3mm}5.51\\
		&	J_4=\hspace{3mm}3.16\\
		&   J_5=\hspace{3mm}2.91\\
		&   J_6=-1.34\\
		&   J_7= -9.68
		\end{aligned}$     & $	\begin{aligned}
		&	J_1=-42.24\\
		&	J_2= -3.71\\
		&	J_3= -0.17\\
		&	J_4=-0.19\\
		&   J_5=-0.15\\
		&   J_6=\hspace{3mm}0.24\\
		&   J_7= -3.71
		\end{aligned}$ & 1654 & 217  \\
		\hline
		Cr$_2$WO$_6$    & $	\begin{aligned}
		&	J_1=  -7.45\\
		&	J_2=\hspace{3mm}2.32\\
		&	J_3= -2.96\\
		&	J_4= \hspace{3mm}5.30\\
		&   J_5= -6.91
		\end{aligned}$     & $	\begin{aligned}
		&	J_1= \hspace{3mm}1.83\\
		&	J_2=\hspace{3mm}2.16\\
		&	J_3=-0.60\\
		&	J_4= \hspace{3mm}1.31\\
		&   J_5= -1.37
		\end{aligned}$ & 121 & 28 \\
		\hline
		NiWO$_4$   & $	\begin{aligned}
		&	J_1= \hspace{3mm}5.67\\
		&	J_2=\hspace{3mm}3.23\\
		&	J_3=\hspace{3mm}1.33\\
		&	J_4=-8.03\\
		&   J_5= -17.94
		\end{aligned}$     & $	\begin{aligned}
		&	J_1=\hspace{3mm}1.95\\
		&	J_2=\hspace{3mm}0.46\\
		&	J_3=\hspace{3mm}0.57\\
		&	J_4=-2.11\\
		&   J_5=-3.97
		\end{aligned}$ & 99 & 21 \\
		\hline
		MnWO$_4$     & $	\begin{aligned}
		&	 J_1= -11.99\\
		&	 J_2=\hspace{3mm}0.01\\
		&	 J_3= -0.97\\
		&	 J_4=-9.29\\
		&    J_5=-20.60
		\end{aligned}$     & $	\begin{aligned}
		&	 J_1=-5.71\\
		&	 J_2=\hspace{3mm}0.04\\
		&	J_3= -0.49\\
		&	J_4= -2.25\\
		&   J_5= -5.20
		\end{aligned}$ & 114 & 20 \\
		\hline
		CoWO$_4$     & $	\begin{aligned}
		&	 J_1=\hspace{3mm}5.96\\
		&	 J_2= -0.32\\
		&	 J_3=\hspace{3mm}2.01\\
		&	 J_4=  -1.71\\
		&    J_5=  -0.66\\
		&    J_6=\hspace{3mm}1.01\\
		&    J_7=  -1.25\\
		&    J_8= -0.64\\
		&    J_9=-0.74
		\end{aligned}$     & $	\begin{aligned}
		&	 J_1=\hspace{3mm}0.68\\
		&	 J_2= -0.13\\
		&	 J_3= -0.07\\
		&	 J_4=  -0.49\\
		&    J_5= -0.48\\
		&    J_6=\hspace{3mm}0.04\\
		&    J_7= -0.48\\
		&    J_8= -0.21\\
		&    J_9=  -0.12
		\end{aligned}$ & 52 & 17 \\
		\hline
	\end{tabular}
\end{table}
\begin{table}[H]
	\centering
	\caption{ Hesineberg exchanges  and transition temperatures in GGA and GGA+$U$ approximation.}
	\label{tab:Hes4}
	\begin{tabular}{lcccc}
		\hline
		Sample     & \makecell{$J_{GGA}$\\ (meV)}   & \makecell{$J_{GGA+U}$\\ (meV)}  & \makecell{$T_{{GGA}}$\\ (K)} & \makecell{$T_{{GGA+U}}$\\ (K)}  \\
		\hline
		Li$_2$MnO$_3$    & $	\begin{aligned}
		&	 J_1=-4.85\\
		&	 J_2=-3.92\\
		&	 J_3=\hspace{3mm}0.69\\
		&	 J_4=\hspace{3mm}0.52\\
		&    J_5=-0.07\\
		&    J_6=-0.03\\
		&    J_7=-0.93\\
		&    J_8=-1.09\\
		&    J_9=-0.74
		\end{aligned}$     & $	\begin{aligned}
		&	 J_1= \hspace{3mm}10.84\\
		&	 J_2=\hspace{3mm}8.12\\
		&	 J_3= \hspace{3mm}1.36\\
		&	 J_4= \hspace{3mm}1.47\\
		&    J_5= \hspace{3mm}0.24\\
		&    J_6=\hspace{3mm}0.28\\
		&    J_7=\hspace{3mm}0.05\\
		&    J_8=-0.20\\
		&    J_9= -0.45
		\end{aligned}$ & 58 & 81 \\
		\hline
		MnTe    & $	\begin{aligned}
		&	 J_1=-32.42\\
		&	 J_2= \hspace{3mm}0.54\\
		&	 J_3=-9.90\\
		&	 J_4=-0.88\\
		&    J_5=-1.25\\
		&    J_6=-0.89\\
		&    J_7=-0.07
		\end{aligned}$     & $	\begin{aligned}
		&	 J_1=-21.04\\
		&	 J_2= \hspace{3mm}0.07\\
		&	 J_3=-2.54\\
		&	 J_4= -0.36\\
		&    J_5=-0.03\\
		&    J_6= -0.51\\
		&    J_7= -0.36
		\end{aligned}$ & 503 & 189 \\
		\hline
	\end{tabular}
\end{table}


\end{document}